\documentstyle[11pt]{article}
\begin{document}

\def\bef{\begin{figure}}
\def\eef{\end{figure}}
\newcommand{\be}[1]{\begin{equation}\label{#1}}
\newcommand{\beq}{\begin{equation}}
\newcommand{\ee}{\end{equation}}
\newcommand{\beqn}[1]{\begin{eqnarray}\label{#1}}
\newcommand{\eeqn}{\end{eqnarray}}
\newcommand{\bd}{\begin{displaymath}}
\newcommand{\ed}{\end{displaymath}}
\newcommand{\mat}[4]{\left(\begin{array}{cc}{#1}&{#2}\\{#3}&{#4}
\end{array}\right)}
\newcommand{\matr}[9]{\left(\begin{array}{ccc}{#1}&{#2}&{#3}\\
{#4}&{#5}&{#6}\\{#7}&{#8}&{#9}\end{array}\right)}
\def\lsim{\raise0.3ex\hbox{$\;<$\kern-0.75em\raise-1.1ex
\hbox{$\sim\;$}}}
\def\gsim{\raise0.3ex\hbox{$\;>$\kern-0.75em\raise-1.1ex
\hbox{$\sim\;$}}}
\def\abs#1{\left| #1\right|}
\def\simlt{\mathrel{\lower2.5pt\vbox{\lineskip=0pt\baselineskip=0pt
           \hbox{$<$}\hbox{$\sim$}}}}
\def\simgt{\mathrel{\lower2.5pt\vbox{\lineskip=0pt\baselineskip=0pt
           \hbox{$>$}\hbox{$\sim$}}}}
\def\unity{{\hbox{1\kern-.8mm l}}}
\def\epr{E^\prime}
\newcommand{\al}{\alpha}
\def\16p{16\pi^2}
\newcommand{\eps}{\varepsilon}
\def\ep{\epsilon}
\def\ga{\gamma}
\def\Ga{\Gamma}
\def\om{\omega}
\def\OM{\Omega}
\def\la{\lambda}
\def\La{\Lambda}
\def\al{\alpha}
\newcommand{\ov}{\overline}
\renewcommand{\to}{\rightarrow}
\def\tm{{\widetilde{m}}}
\def\mcirc{{\stackrel{o}{m}}}
\def\dem{\delta m^2} 
\def\sint{\sin^2 2\theta} 
\def\tant{\tan 2\theta} 
\def\tanL{\tan 2\theta^L}
\def\tanR{\tan 2\theta^R}
\newcommand{\tanb}{\tan\beta}
\def\brf{{\bf f}}
\def\bbf{\bar{\bf f}}
\def\bF{{\bf F}}
\def\bbF{\bar{\bf F}}
\def\bFp{{\bf F^\prime}}
\def\bbFp{\bar{\bf F^\prime}}
\def\bY{{\bf Y}}
\def\by{{\bf y}}
\def\bX{{\bf X}}
\def\bM{{\bf M}}
\def\bA{{\bf A}}
\def\bB{{\bf B}}
\def\bG{{\bf G}}
\newcommand{\tphi}{\tilde{\phi}}
\newcommand{\Tphi}{\tilde{\Phi}}
\def\Bsi{{\bar{\Psi}}}
\newcommand{\bx}{\bar{\rm X}} 
\newcommand{\wx}{{\rm X}} 
\newcommand{\bv}{\bar{\rm V}} 
\newcommand{\wv}{{\rm V}} 
\newcommand{\tl}{\tilde{l}} 
\newcommand{\tq}{\tilde{q}}
\newcommand{\tu}{\tilde{u}}
\newcommand{\td}{\tilde{d}}
\newcommand{\tuc}{\tilde{u}_c} 
\newcommand{\tdc}{\tilde{d}_c} 
\newcommand{\tec}{\tilde{e}_c} 
\newcommand{\TQ}{\tilde{Q}} 
\newcommand{\TU}{\tilde{U}}
\newcommand{\TE}{\tilde{E}} 
\newcommand{\TUC}{\tilde{U}_c} 
\newcommand{\TEC}{\tilde{E}_c} 
\newcommand{\TQC}{\tilde{Q}_c} 
%




\begin{titlepage}

\begin{flushright}
{\large hep-ph/9811447 }  \\ 
DFAQ-98/TH/01 \\ 
FISIST/6-98/CFIF \\ 
November 1998 
\end{flushright}

\vspace{2.0cm}

\begin{center}

{\Large \bf 
Grand unified textures for neutrino and quark mixings} \\

\vspace{0.7cm}

{\large Zurab Berezhiani${}^{a,}$\footnote{
E-mail: berezhiani@fe.infn.it }
and Anna Rossi${}^{b,}$\footnote{
Address after November 1, 1998: Universit\`a di Padova and 
INFN Sezione di Padova, I-35131 Padova, Italy. 
E-mail: arossi@pd.infn.it }
}
\vspace{5mm}

{\it ${}^a$ Universit\`a dell'Aquila, 
I-67010 Coppito, L'Aquila, Italy, and \\
Institute of Physics, Georgian Academy of Sciences,
380077 Tbilisi, Georgia}\\

{\it  ${}^b$ CFIF - Instituto Superior Tecnico, 
1096 Lisbon, Portugal}

\end{center}

\vspace{10mm}

\begin{abstract}
\noindent
The atmospheric neutrino data imply large mixing between the 
$\nu_\mu$ and $\nu_\tau$ states, $\theta_{23} = (45\pm 12)^\circ$, 
while the MSW solution to the solar neutrino problem needs 
very small mixing angle $\theta_{12} = (2\pm 1)^\circ$. 
In the quark sector the situation is rather opposite -- 
the 23 mixing is tiny, $\theta_{23}\simeq 2^\circ$, 
versus reasonable 12 mixing, $\theta_{12}\simeq 13^\circ$. 
We show that such complementary patterns of the quark and 
leptonic mixings could naturally emerge in the context of 
the $SU(5)$ grand unification, assuming that the fermion 
mass matrices have the Fritzsch-like structures 
but their off-diagonal entries are not necessarily symmetric. 
Such a picture exhibits a `see-saw' like correspondence 
between the quark and leptonic mixing patterns 
so that the smaller the quark mixing angle is, 
the  larger  the corresponding leptonic mixing angle becomes. 
This fact simply follows from the fermion multiplet structure 
in $SU(5)$. 
We also discuss a model with  horizontal symmetry $U(2)$ 
in which the discussed pattern of the mass matrices can 
emerge rather naturally. 

\end{abstract}
\end{titlepage}


\section{Introduction} 

Signals of  neutrino oscillations accumulated during 
past several years impose strong constraints 
on the mass and mixing pattern of the three known neutrinos 
$\nu_{e,\mu,\tau}$. 
In particular, the atmospheric neutrino (AN) anomaly, 
a long-standing discrepancy by almost a factor of 2 
between the predicted and observed $\nu_\mu/\nu_e$ ratio
of the atmospheric neutrino fluxes,  has recently received
a strong confirmation from a high statistics experiment by the 
SuperKamiokande Collaboration \cite{ANP}. These data     
indicate that the zenith angle/energy dependence 
of the atmospheric $\nu_\mu$ flux 
is compatible with $\nu_\mu-\nu_\tau$ oscillation 
within the following parameter range: 
\beqn{AN}
&&
\delta m^2_{\rm atm}= (0.5-6)\times 10^{-3}~ {\rm eV}^2,
\nonumber \\ 
&&
\sin^2 2\theta_{\rm atm}>0.82 
\eeqn     
(the best-fit values are 
$\dem_{\rm atm}=2.2\times 10^{-3}~\mbox{eV}^2$ and  
$\sint_{\rm atm}=1.0$),  
and disfavour the $\nu_\mu -\nu_e$ oscillation as a 
dominant reason for the AN  anomaly. 

On the other hand, the solar neutrino (SN) problem, 
an energy dependent deficit of the solar $\nu_e$ fluxes 
indicated by the solar neutrino experiments cannot 
be explained by nuclear or astrophysical reasons \cite{SNP}. 
The most natural solution is provided by the resonant MSW 
oscillation \cite{MSW} of $\nu_e$ into $\nu_\mu$, $\nu_\tau$   
or their mixture, which requires the following parameter 
range \cite{BKS}:   
\beqn{SN}
&&
\delta m^2_{\rm sol}= (3-10)\times 10^{-6}~{\rm eV}^2,
\nonumber \\ 
&& 
\sin^2 2\theta_{\rm sol}=(0.1-1.5)\cdot 10^{-2} 
\eeqn     
(the best-fit values are 
$\dem_{\rm sol}=5\times 10^{-6}~\mbox{eV}^2$ 
and  $\sint_{\rm sol}=6\times 10^{-3}$).\footnote{
The long wavelength {\it Just-so} 
oscillation from the Sun to the Earth \cite{just-so} provides 
as good a fit as that of MSW,  with the parameter range 
$\delta m^2_{\rm sol}\sim 10^{-10}$ eV$^2$ and 
$\sin^2 2\theta_{\rm sol}\sim 1$. However, in this 
paper we mainly concentrate on the pattern dictated by the 
MSW solution. 
}

The explanation of the fermion mass and mixing pattern  
is beyond the capacities of the standard model (SM) and the 
neutrino case represents a part of the flavour problem.  
The masses of the charged fermions $q_i=(u_i,d_i)$, $u^c_i$, $d^c_i$; 
$l_i=(\nu_i,e_i)$, $e^c_i$  ($i=1,2,3$ is a family index)  
emerge from the Yukawa terms: 
\be{Yuk}
\phi_2 u^c_i \bY_u^{ij} q_j 
+ \phi_1 d^c_i \bY_d^{ij} q_j + \phi_1 e^c_i \bY_e^{ij} l_j 
\ee
where $\phi_{1,2}$ are the Higgs doublets: 
$\langle \phi_{1,2} \rangle = v_{1,2}$, 
$(v_1^2 + v_2^2)^{1/2} =v_w=174$ GeV
and $\bY_{u,d,e}$ are general complex $3\times 3$ matrices 
of the coupling constants.\footnote{
Rather spontaneously we have chosen our notations as the 
ones adopted in the supersymmetric SM which in the following 
will be mentioned simply as SM, 
while the ordinary standard model, if any,  
we shall recall as a non-supersymmetric SM.
Barring numerical details, all main arguments put 
forward in our paper will be valid 
also for non-supersymmetric case.  
}  
In order to identify the physical basis of the fermion 
mass eigenstates,  the Yukawa matrices $\bY_{u,d,e}$ 
and thus the fermion mass matrices 
should be brought to the diagonal form 
via the bi-unitary transformations: 
\beqn{diag}
&&
U'^T_{u}\bY_u U_{u}=
\bY_u^{D}=\mbox{Diag}(Y_u,\,Y_c,\,Y_t)
\nonumber \\
&&
U'^T_{d}\bY_d U_{d}=
\bY_d^{D}=\mbox{Diag}(Y_d,\,Y_s,\,Y_b) 
\nonumber \\ 
&&
U'^T_{e}\bY_e U_{e}=
\bY_e^{D}=\mbox{Diag}(Y_e,\,Y_\mu,\,Y_\tau)
\eeqn
Hence, the quark mass eigenstates mix in the charged current 
$\bar{u}_i\ga^\mu (1+\ga_5)V_q^{ij} d_j$,  
and the Cabibbo-Kobayashi-Maskawa (CKM) matrix 
$V_q=U^\dagger_{u}U_{d}$ can be parametrized as \cite{maiani}:
\be{CKM}
V_q = \matr{V_{ud}}{V_{us}}{V_{ub}} {V_{cd}}{V_{cs}}{V_{cb}} 
{V_{td}}{V_{ts}}{V_{tb}} = 
\matr{c_{12}c_{13}}{s_{12}c_{13}}{s_{13}e^{-i\delta}}
{-s_{12}c_{23}-c_{12}s_{23}s_{13}e^{i\delta}}
{c_{12}c_{23}-s_{12}s_{23}s_{13}e^{i\delta}} {s_{23}c_{13}} 
{s_{12}s_{23}-c_{12}c_{23}s_{13}e^{i\delta}}
{-c_{12}s_{23}-s_{12}c_{23}s_{13}e^{i\delta}} {c_{23}c_{13}} 
\ee
where $s_{ij}$ and $c_{ij}$ respectively stand for the sines 
and cosines of three mixing angles $\theta_{12}$, $\theta_{23}$
and $\theta_{13}$, and $\delta$ is the CP-violating phase. 

As for neutrino masses, they emerge only from the higher 
order effective operator \cite{BEG}: 
\be{Yuk-nu}
\frac{\phi_2\phi_2}{M}\, l_i \bY_\nu^{ij} l_j \,, ~~~~~~~ 
\bY_\nu^{ij}=\bY_\nu^{ji} 
\ee
where $M\gg v_w$ is some cutoff scale and  
$\bY_\nu$ is the (symmetric) matrix of the dimensionless coupling 
constants. Thus, while the charged fermion masses are linear with 
respect to the weak scale $v_{w}$, the 
neutrino masses are bilinear which makes 
magnitudes of the latter naturally small.\footnote{Any 
known mechanism for the neutrino masses effectively reduces 
to the operator (\ref{Yuk-nu}) after integrating out the 
relevant heavy states (e.g. in the `seesaw' mechanism, after 
integrating out the right-handed Majorana neutrinos 
with masses $\sim M$).
}
One can go to the neutrino physical basis $\nu_{1,2,3}$ 
by the unitary transformation 
\be{diag-nu}
U_{\nu}^T\bY_\nu U_{\nu}=\bY_\nu^{D} = 
\mbox{Diag}(Y_1,\,Y_2,\,Y_3)
\ee
and the mixing matrix $\tilde{V}_l=U_{e}^\dagger U_{\nu}$ 
in the leptonic current 
$\bar{e}_i\ga^\mu(1+\ga_5)V_l^{ij}\nu_j$ 
can be presented as: 
\be{MNS}  
\tilde{V}_l = V_l P_\nu = 
\matr{V_{e1}}{V_{e2}}{V_{e3}} {V_{\mu1}}{V_{\mu2}}{V_{\mu3}} 
{V_{\tau1}}{V_{\tau2}}{V_{\tau3}} 
\matr{1}{0}{0} {0}{e^{i\delta_2}}{0} {0}{0}{e^{i\delta_3}} 
\ee 
where the first factor $V_l$ 
can be parametrized in a manner similar to (\ref{CKM}).  
It relates the neutrino flavour eigenstates 
$(\nu_e,\nu_\mu,\nu_\tau)$ to 
the mass eigenstates $(\nu_1,\nu_2,\nu_3)$, and describes   
the neutrino oscillation phenomena. 
(Due to the Majorana nature of neutrinos, the matrix 
$\tilde{V}_l$ contains two additional phases $\delta_{2,3}$, 
but these are not relevant for neutrino oscillations).  
In the following, we distinguish 
the quark and lepton mixing angles in $V_q$ and $V_l$ 
 by the subscripts `$q$' and `$l$', respectively.

The mass spectrum of the quarks and charged leptons is spread over
five orders of magnitude, from MeVs to hundreds of GeVs, 
with a strong inter-family hierarchy \cite{PDG}:\footnote{
Following the tradition, for the heavy quarks $t,b,c$,
we refer to their running masses respectively at $\mu=m_{t,b,c}$,
and for the light quarks $u,d,s$ -- at $\mu=1$ GeV. 
The quoted value of $m_t$ corresponds to the `pole' mass  
$M_t=173.8\pm 5.2$ GeV \cite{PDG}.   
} 
\beqn{masses}
&&
m_t=163\pm 5 ~{\rm GeV}, ~~~~
m_c=1.1-1.4 ~{\rm GeV}, ~~~~~
m_u=2-7 ~{\rm MeV}    \nonumber \\
&&
m_b=4.1-4.4 ~{\rm GeV}, ~~~
m_s= 80-230 ~ {\rm MeV}, ~~~
m_d=4-12 ~{\rm MeV}  \nonumber \\
&&
m_\tau=1.777 ~{\rm GeV}, ~~~~~~~
m_\mu=105.7 ~{\rm MeV}, ~~~~~~~~
m_e=0.511 ~{\rm MeV}
\eeqn
and the quark mixing angles are known experimentally with a very 
good accuracy:\footnote{
We find it instructive to present the values of mixing angles 
in degrees, which correspond to the CKM matrix elements   
$|V_{us}|=0.221\pm 0.002$, $|V_{cb}|=0.040\pm 0.004$ 
and $|V_{ub}/V_{cb}|=0.08\pm 0.02$  \cite{PDG}. 
} 
\be{q-ang}
\theta_{23}^q = (2.3 \pm 0.2)^\circ, ~~~~~ 
\theta_{12}^q = (12.7 \pm 0.1)^\circ, ~~~~~ 
\theta_{13}^q = (0.18 \pm 0.04)^\circ 
\ee

As for the neutrinos, information on their masses 
can be extracted directly from the ranges of 
$\dem_{\rm atm}$ and $\dem_{\rm sol}$
needed for the explanation of the AN and SN anomalies. 
Barring the less natural possibility 
that the neutrino mass eigenstates $\nu_{1,2,3}$ 
are strongly degenerate and 
assuming the mass hierarchy $m_3 >m_2> m_1$,  
these data translate into the following values 
for neutrino masses:
\be{nu-spectr} 
m_3 = (4.7_{-2.5}^{+3.0})\times 10^{-2} ~ {\rm eV},  ~~~~~ 
m_2 = (2.2_{-0.5}^{+1.0})\times 10^{-3} ~ {\rm eV}. 
\ee
Hence, the neutrino mass hierarchy  $m_3/m_2 \sim 10-50$ 
is similar to that of the charged leptons or down quarks.   
However, the magnitudes of the neutrino mixing 
angles\footnote{
As far as the $V_{e3}$ element is concerned, 
for $\dem_{\rm atm} > 2\times 10^{-3}$ eV$^2$ 
the limit  $\theta^l_{13}<13^\circ$ follows 
from the CHOOZ experiment \cite{CHOOZ}. But also  
for the $\dem_{\rm atm}$ region not covered by CHOOZ,    
the recent fits disfavour $\theta^l_{13}$ larger than 
$20^\circ$ or so \cite{Lisi}. In a whole, 
$\theta^l_{13}\approx 0$ provides the best data fit 
both for AN and SN cases \cite{BHSSW}. 
}  
\be{nu-ang}
\theta_{23}^l= (45\pm 12.5)^\circ , ~~~~~ 
\theta_{12}^l= (2.2\pm 1.3)^\circ , ~~~~~
\theta_{13}^l < (13-20)^\circ , 
\ee 
are in a dramatic contrast with the corresponding angles 
(\ref{q-ang}) in the quark mixing.   
In other words, the AN anomaly points to  maximal 23 mixing 
in leptonic sector versus very small 23 mixing of quarks, 
and on the contrary, the MSW solution implies a very small 12 
lepton mixing angle  versus the reasonably large value 
of the Cabibbo angle.

As said above, the SM does not contain any theoretical input 
restricting the structure of the matrices 
$\bY_{u,d,e}$ and $\bY_\nu$ and thus 
fermion mass hierarchy and mixing pattern remain unexplained. 
Concerning the neutrinos, also the mass scale $M$ remains  a free 
parameter. One can only conclude that if the maximal 
constant in $\bY_\nu$ is order of the top Yukawa constant,  
$Y_3\sim Y_t\sim 1$, then the mass value $m_3$ in 
(\ref{nu-spectr}) points to the scale $M\sim 10^{15}$ GeV, 
rather close to the grand unified scale.  
On the other hand, the drastic difference of the neutrino 
mixing pattern from that of the quarks at first glance suggests 
that the neutrino mass texture is very special and 
it indicates no similarity to that of the quarks 
and charged leptons. 
During the past years many models have been produced 
suggesting various exotic input textures 
for understanding the neutrino mixing pattern 
(e.g. refs. \cite{all}; for a more generic discussions 
see refs. \cite{BHSSW,BHS}. 
There have been also attempts to 
obtain the desired pattern in the context 
of grand unification \cite{tavzur,tavzur1,U2-nu}).

In this paper we show that there is a 
{\it simple} and {\it coherent}  way of 
understanding  both the quark and 
neutrino mixing patterns within an unified framework. 
Our consideration is motivated by the following points. 
It is tempting to think that the intriguing empiric relations 
between the masses and mixing angles, 
such as the well-known formula for the Cabibbo angle 
$\,s_{12}^q=\sqrt{m_d/m_s}\,$, 
are not accidental and  the fermion flavour structure 
is intrinsically connected to the peculiarities of some 
underlying theory which fully determines, or at least 
somehow constrains the form of the Yukawa matrices. 
The ideas of supersymmetry, grand unification and horizontal 
symmetry may constitute the essential ingredients of 
flavour physics and can be regarded as the present 
{\em Modus Operandi} for  predictive model building 
(for a review on the fermion mass models see e.g. 
\cite{ICTP} and references therein).

In particular, relations 
between the fermion masses and mixing angles can be obtained
by considering  Yukawa matrix textures with reduced number 
of free parameters, putting  certain elements  to zero. 
For example, one can consider the popular texture 
suggested by Fritzsch \cite{Fritzsch},  
which implies that the fermion mass generation starts from the
$3^{rd}$ family 
and proceeds to lighter families through the mixing terms:   
\be{Fr}
\bY_{u,d,e}=\,\matr{0}{A'_{u,d,e}}{0}
{A_{u,d,e}}{0}{B'_{u,d,e}}{0}{B_{u,d,e}}{C_{u,d,e}},
\ee 
where all elements are generically  complex and obey 
the additional condition:  
\be{A-B}
|A'_f|=|A_f|, ~~~~~~ |B'_f|=|B_f|; ~~~~~~~~~ f=u,d,e 
\ee
This texture could emerge due to horizontal symmetry reasons, 
and the "symmetricity" property (\ref{A-B}) 
can be motivated in the context of the left-right symmetric 
models \cite{Fritzsch}, or of the $U(3)_H$ horizontal 
symmetry \cite{PLB85} (see also the 
discussion in ref. \cite{BLM}).

This pattern has many striking properties. 
For example, it nicely links 
the observed value of the Cabibbo angle, 
$V_{us}\approx \sqrt{m_d/m_s}$,  
to the observed size of the CP-violation in the $K-\bar{K}$ 
system, moreover that the predicted magnitude 
$|V_{ub}/V_{cb}|\approx \sqrt{m_u/m_c}$ is also in 
good agreement with experiment.  
Unfortunately, the Fritzsch texture contains a strong 
conflict between the small value of  $V_{cb}$ and 
the large top mass and there is no parameter space in which 
these observables could be reconciled \cite{KFP}.  

However, one can consider  textures like (\ref{Fr}) 
without the symmetricity in the 23 block of $\bY_d$, 
$B'_{d}\neq B_{d}$. 
Then one could achieve  the small 
magnitude of $V_{cb}$ at the price of taking the parameter 
$b_d=B'_d/B_d$ considerably larger than 1.\footnote{
A particular texture with the maximal asymmetry, $B'_d=C_d$, 
was suggested in ref. \cite{Branco}, and its relevance 
for small quark mixing angles was pointed out. 
The fact that in grand unified theories a similar asymmetry  
could lead to the large neutrino mixing was demonstrated 
in the $SO(10)$ based models \cite{tavzur1}.}  
In the present paper we present a simple observation:  
in the context of the $SU(5)$ grand unified theory 
such a choice naturally implies that the 23 mixing in the 
leptonic sector becomes large, and  the parameter 
$b_e=B_e/B'_e$ increases in parallel with $b_d$   
since in SU(5) the Yukawa matrices are related 
as $\bY_e = \bY_d^T$, modulo certain Clebsch factors. 
Our point can be simply expressed as follows.
If the down-quark and charged-lepton matrices 
had the symmetric Fritzsch texture, then we would have 
$\tan\theta_{23}^d= (m_s/m_b)^{1/2}$ and 
$\tan\theta_{23}^e = (m_\mu/m_\tau)^{1/2}$,   
which are both unsatisfactory: the former is too big as compared 
to $|V_{cb}|$, while the latter is too small for $|V_{\mu3}|$. 
On the other hand, whenever 
the symmetricity condition is abandoned, 
non of these angles can be predicted in terms of mass 
ratios since their values now depend on the amount of asymmetry 
between the 23 and 32 entries, i.e. on the  factors $b_d$ and $b_e$.  
However, in the context of the SU(5) theory  there emerges 
the following product rule: 
\be{rule23} 
\tan\theta_{23}^d \tan\theta_{23}^e \sim 
\left(\frac{m_\mu m_s}{m_\tau m_b}\right)^{1/2}. 
\ee
Therefore, if $\tan\theta_{23}^d$ becomes smaller than 
$(m_s/m_b)^{1/2}$, then  $\tan\theta_{23}^e$ should 
correspondingly increase over $(m_\mu/m_\tau)^{1/2}$, 
and when the former reaches the value $|V_{cb}|\simeq 0.05$, 
the latter becomes $\sim 1$  
(this happens for $b_{d,e}\sim 8$). 
Though these estimates are not precise (the exact expressions 
will be given in section 2), they qualitatively rather well 
demonstrate the "seesaw" like correspondence between the 
quark and lepton mixing angles whenewer their magnitudes are 
dominated by the rotation angles coming from the 
down fermions.
A similar argument can be applied also to the 12 mixing: 
\be{rule12}  
\tan\theta_{12}^d \tan\theta_{12}^e \sim 
\left(\frac{m_e m_d}{m_\mu m_s}\right)^{1/2},  
\ee
The relation $V_{us}\simeq (m_d/m_s)^{1/2}$ 
points that the 12 block of $\bY_d$ should be nearly symmetric, 
and hence we expect that $V_{e2}\sim (m_e/m_\mu)^{1/2}$.   
All of these will be discussed in details in the next section. 

In principle, reasonable contributions to the mixing angles 
will emerge if also the upper quark and neutrino 
couplings,  $\bY_u$ and $\bY_\nu$, have (symmetric) textures 
like (\ref{Fr}). In this case the proper fit of the quark 
and lepton mixing patterns can be achieved for  rather 
moderate asymmetry, $b_{d,e} \geq 2$. 
This possibility is  discussed in section 3, where we also 
consider models with a horizontal symmetry which 
could provide the natural realization of the suggested 
texture. Finally, at the end we briefly outline 
our results and their implications.

\section{Fritzsch-like $\bY_{d,e}$ and  
diagonal  $\bY_{u,\nu}$ in SU(5) }

In the $SU(5)$ model the quark and lepton states of each family 
fit into the following multiplets: 
$\bar{5}_i=(d^c, l)_i$, ~  $10_i =(u^c, e^c, q)_i$  
(here and in the following the $SU(5)$ indices are suppressed, 
and $i=1,2,3$ is a family index).  
As for the Higgs doublets $\phi_{1,2}$, together with their 
colour triplet partners $T,\bar{T}$ they form the 
representations 
$H=(T,\phi_2)\sim 5$ and $\bar H=(\bar T,\phi_1)\sim \bar5$.  
The theory contains also the chiral superfield 
in the adjoint representation, $\Phi \sim 24$, 
which at the scale $M_G\sim 10^{16}$ GeV breaks the $SU(5)$ 
symmetry down to $SU(3)\times SU(2)\times U(1)$.

The Yukawa terms responsible for the 
fermion masses are the following: 
\be{Yuk-su5}
\bar{H} 10_i\bG^{ij}\bar5_j + H 10_i\bG_u^{ij}10_j + 
\frac{HH}{M}\bar5_i\bG_\nu^{ij}\bar5_j
\ee
where the Yukawa constant matrices $\bG_{u}$ and $\bG_{\nu}$ 
are symmetric due to $SU(5)$ symmetry reasons 
while the form of $\bG$ is not constrained. 
After the $SU(5)$ symmetry breaking 
these terms reduce to the SM Yukawa couplings (\ref{Yuk}) with 
\be{relat}
\bY_e=\bG, ~~~~ \bY_d=\bG^T, ~~~~ \bY_u=\bG_u, ~~~~ 
\bY_\nu=\bG_\nu
\ee
Without loss of generality, the matrices $\bG_u$ and $\bG_\nu$ 
can be taken diagonal,   
so that the weak mixing 
matrices in both the quark and leptonic sectors is determined 
by the unitary matrices rotating the left states of the 
down fermions: $V_q=U_d$ and $V_l=U_e^\dagger$. 
On the other hand, 
since $\bY_d=\bY_e^T$, 
we get that $U_{d}=U'_{e}$ and $U_{e}=U'_{d}$, 
so that the rotation angles of the left down quarks 
(charged leptons) are related to the unphysical angles 
rotating the right states of the charged leptons 
(down quarks).\footnote{ 
The latter have no physical 
significance for the low energy theory (the SM) and could be 
relevant only for the baryon number violating processes 
mediated by the $SU(5)$ gauge fields or Higgses/Higgsinos. 
}
Therefore, the smallness of the quark mixing angle 
$\theta^q_{23}$ does not necessarily imply the 
smallness of the leptonic mixing $\theta^l_{23}$   
but rather the opposite, 
it can point to the large value of the latter.

Indeed, imagine that in the basis where $\bG_u$ and $\bG_\nu$ 
are diagonal: 
\be{u-nu}
\bG_u=\bY_u^D=\matr{Y_u}{0}{0} {0}{Y_c}{0} {0}{0}{Y_t}, ~~~~~~~
\bG_\nu=\bY_\nu^D=\matr{Y_1}{0}{0} {0}{Y_2}{0} {0}{0}{Y_3}, 
\ee
the matrix $\bG$ has a Fritzsch-like form: 
\be{bG-form}
\bG = \matr{0}{G_{12}}{0} {G_{21}}{0}{G_{23}} {0}{G_{32}}{G_{33}} 
\ee
In this case, as we have already remarked, the 
large value of the leptonic rotation angle is 
related to the smallness of the corresponding quark angle. 
However, if the entries of the matrix $\bG$ are just constants,  
we would face the well-known 
problem of the down-quark and charged-lepton degeneracy 
in the $SU(5)$ symmetry limit: since $\bY_d^D=\bY_e^D$, then 
we have $Y_{d,s,b}=Y_{e,\mu,\tau}$. 
Nevertheless, the $Y_b=Y_\tau$ unification is a definite success of 
the SUSY $SU(5)$ GUT. 
After accounting for the renormalization  running of the 
Yukawa constants from $M_G$ to lower-energy scales, 
it provides a nice explanation of the magnitude of the bottom 
mass and its intimate relation to the large top mass.  
However, the other predictions 
$Y_s=Y_\mu$ and $Y_d=Y_e$ are wrong: they imply 
$m_s/m_d=m_\mu/m_e\simeq 200$,  an order of magnitude in conflict 
with the current algebra estimate $m_s/m_d\simeq 20$.

It is natural, however, to consider that the $\bG^{ij}$ 
are  in fact operators dependent on the adjoint superfield of $SU(5)$:  
$G_{ij} = G_{ij}(\Phi)$. The latter should be understood as 
expansion series  $G_{ij}(\Phi)= 
G^{(0)}_{ij} + G^{(1)}_{ij}\frac{\Phi}{M_s} + ...$, 
where $M_s$ is some fundamental scale larger than $M_G$ 
(it can be e.g. the string scale or the scale where some GUT 
with larger symmetry reduces to $SU(5)$).  
In other words, one can assume that the 
couplings $\bar{H}10_i\bG^{ij}\bar5_j$ contain the effective 
higher-order operators 
$\frac{\Phi}{M_s}\bar{H}10_iG_{ij}^{(1)}\bar5_j$ etc., just 
on the same footing as the last term in (\ref{Yuk-su5}). 
Since in general the operator $\Phi\cdot H$ is represented 
by the tensor product $24\times \bar5=\bar5 + \ov{45}$,  
it can distinguish the corresponding 
entries in the matrices $\bY_e$ and $\bY_d$.
Needless to say that in the field theory context such 
operators can be effectively induced from the renormalizable 
Lagrangian, by integrating out the additional superheavy states 
with masses $\sim M_s$ 
(so called Frogatt-Nielsen or {\it universal seesaw} 
mechanism \cite{FN}), 
much in the same way as the effective neutrino operator is 
obtained in the context of the seesaw mechanism 
by integrating out the heavy right-handed Majorana neutrinos.    
In addition, the ratio $\eps = \langle \Phi\rangle/M_s $
can be used as a small parameter for understanding the 
fermion mass hierarchy.

Therefore, for the charged-lepton and the down-quark 
Yukawa matrices we get: 
\be{Fr-e,d}
\bY_e = \matr{0}{A'_{e}}{0} {A_{e}}{0}{B'_{e}} {0}{B_{e}}{C_e}, 
~~~~~~ 
\bY_d = \matr{0}{A'_{d}}{0} {A_{d}}{0}{B'_{d}} {0}{B_{d}}{C_d}, 
\ee
With a proper redefinition of the fermion phases 
all elements in (\ref{Fr-e,d}) can be made real and positive. 

As we said above, generically the constants $G_{ij}$ can be 
functions of the adjoint superfield $\Phi$, 
so that the tensor product $G_{ij}(\Phi) \bar{H}$ can contain  
both $\bar5 + \ov{45}$ channels and therefore the corresponding 
entries between $\bY_e$ and $\bY_d^T$ should be distinguished 
by some Clebsch coefficients of $SU(5)$.  
However, for simplifying our analysis, and for designing 
more predictive ansatz, we impose some rather natural constraints. 
In particular, inspired by the success of $Y_b\simeq Y_\tau$ 
unification, we assume that $G_{33}$ is a $SU(5)$ singlet, 
so that $C_d=C_e=G_{33}$. Analogously, following another 
interesting relation $Y_dY_s\simeq Y_eY_\mu$ we assume that also 
$G_{12}$ and $G_{21}$ are $SU(5)$ singlets. Furthermore, 
motivated by the celebrated formula for the Cabibbo angle  
$s^q_{12}\simeq \sqrt{m_d/m_s}$, we suppose that 
the matrix $\bG$ is symmetric in the 12 block, $G_{12}=G_{21}$. 
All these imply that 
\beqn{restrict} 
&&
C_d=C_e ~(=C),  \nonumber \\ 
&&
A_d=A'_d=A'_e=A_e ~(=A).  
\eeqn 
Thus only the $G_{23}(\Phi)$ and $G_{32}(\Phi)$ 
entries are left unconstrained.  
They contain nontrivial $SU(5)$ Clebsches 
breaking the quark and lepton symmetry: 
\be{Clebsch}
B'_d = k'B_e,  ~~~~~~ B_d = k B'_e 
\ee 
where the coefficients $k$ and $k'$ are not necessarily 
the same, since generically the tensor products 
$G_{23}(\Phi)\cdot \bar{H}$ and $G_{32}(\Phi)\cdot \bar{H}$
can emerge in different combinations of $\bar5$ and $\ov{45}$. 
In addition, these entries are not necessarily symmetric, 
and we introduce the asymmetry parameters $b_e=B_e/B'_e$ 
and $b_d=B'_d/B_d = \frac{k'}{k}b_e$.   
Then, identifying $B_e=B$ and $b_e=b$, the matrices 
$\bY_e$ and $\bY_d$ can be represented as: 
\be{Fred}
\bY_e = \matr{0}{A}{0} {A}{0}{\frac{1}{b}B} {0}{B}{C}, 
~~~~~~ 
\bY_d = \matr{0}{A}{0} {A}{0}{k'B} {0}{\frac{k}{b}B}{C}, 
\ee
Therefore, we have an ansatz depending on six parameters,   
three Yukawa entries $A,B,C$ and three Clebsch factors 
$k,k'$ and $b$. 
As far as $\bY_u$ and $\bY_\nu$ are taken diagonal, 
the patterns of the quark and lepton mixings are 
completely determined by the form of $\bY_{d,e}$. 
The Yukawa eigenvalues $Y_{e,\mu,\tau}$ and $Y_{d,s,b}$   
as well as the mixing angles 
$s^q_{12},s^q_{23},s^q_{13}$ and 
$s^l_{12},s^l_{23},s^l_{13}$ can be expressed 
in terms of the parameters in (\ref{Fred}) and 
hence at the GUT scale there should emerge six relations 
between these physical quantities.  


The GUT scale Yukawa constants are linked to the physical 
fermion masses (\ref{masses}) through the 
renormalization group equations (RGE). 
For moderate values of $\tanb=v_2/v_1$, one obtains at 
one-loop (see e.g. \cite{RG}):
\beqn{RG}
&& 
m_u=Y_uR_u\eta_u B_t^3 v_2\,,~~~~
m_d=Y_dR_d\eta_d v_1\,,~~~~
 m_e=Y_e R_e v_1  \nonumber  \\
&&
m_c=Y_cR_u\eta_c B_t^3 v_2\,,~~~~~
m_s=Y_sR_d\eta_s  v_1\,,~~~~
m_{\mu}=Y_{\mu} R_e v_1 \\
&&
m_t=Y_t R_u B_t^6 v_2\,,~~~~~~
m_b=Y_bR_d\eta_b B_t v_1\,,~~~~
m_{\tau}=Y_{\tau} R_e v_1 \nonumber
\eeqn
where the factors $R_{u,d,e}$ account for the gauge-coupling 
induced running from the GUT scale $M_G\simeq 10^{16}$ GeV 
to the SUSY breaking scale $M_S\simeq M_t$, and 
the factors $\eta_f$ encapsulate the QCD+QED running from 
$M_S$ down to $m_f$ for $f=b,c$ 
(or to $\mu=1$ GeV for the light quarks $f=u,d,s$). 
Namely, for $\al_s(M_Z)=0.118\pm 0.005$ we have 
\beqn{etas}
&&
R_u=3.33\pm 0.07, ~~~~~ R_d=3.25\pm 0.07, 
~~~~~ R_e=1.49   \nonumber \\ 
&&
\eta_b=1.52\pm 0.04, ~~~~~ \eta_c=2.02^{+0.16}_{-0.13}, 
~~~~~ \eta_{u,d,s}=2.33^{+0.29}_{-0.21} 
\eeqn 
The factor $B_t$ includes the running induced by the 
large top quark Yukawa constant ($Y_t\sim 1$):
\be{B-t}
B_t=\exp\left[-\frac{1}{16\pi^2}\int_{\ln M_S}^{\ln M_G}
Y_t^2(\mu) {\mbox d}(\ln\mu)\right]
\ee
$B_t$ as a function of the GUT scale value $Y_t$ is shown  
in Fig. 1. We see that for $Y_t$ varying from a lower
limit $Y_t=0.5$ imposed by the top pole-mass value to 
a perturbativity limit $Y_t\approx 3$, the factor 
$B_t$ decreases from $0.9$ to $0.7$. 
As for the CKM mixing angles, their physical 
values are related to their values at the GUT scale 
(labelled by superscript G) as follows:\footnote{ 
We shall not take into account 
the analogous RGEs \cite{RG-nu} 
for the neutrino masses and mixing,  
since the experimental data (\ref{nu-spectr}) and 
(\ref{nu-ang}) still contain big error bars 
and such an improvement is not justified.  
In addition, the RGE effects for the neutrino sector 
are model dependent:  
the renormalization of the constants $\bY_\nu$ 
depends on the concrete features of the underlying model 
effectively inducing  the $d=5$ operator (\ref{Yuk-nu}).  
} 
\be{RG-CKM}
s_{12}^q=s_{12}^{qG}, ~~~~~~ s_{23}^q=s_{23}^{qG}B_t^{-1}, 
~~~~~~ s_{13}^q=s_{13}^{qG}B_t^{-1}. 
\ee

Let us now discuss these predictions.  
Using the formulas given in the Appendix, 
we readily obtain the modified version of the $b-\tau$ 
Yukawa unification at the GUT scale:
\be{b-tau}
C=
Y_\tau\left[1-(b_e+b_e^{-1})\frac{Y_\mu-Y_e}{Y_\tau}\right]^{1/2} 
=Y_b\left[1-(b_d+b_d^{-1})\frac{Y_s-Y_d}{Y_b}\right]^{1/2} 
\ee 
and also the following relations:  
\be{dsb}
A^2C = Y_eY_\mu Y_\tau = Y_dY_sY_b ,  
\ee
\be{sb}
B_eB'_e=\frac{1}{b}B^2=(Y_\mu-Y_e)Y_\tau , ~~~~~  
B_dB'_d=\frac{kk'}{b}B^2=(Y_s-Y_d)Y_b . 
\ee
In the following we directly substitute the Yukawa constant 
ratios with the corresponding mass ratios 
when the latter are RGE invariant (c.f. (\ref{RG})), 
e.g. $Y_\mu/Y_\tau=m_\mu/m_\tau$, $Y_d/Y_s=m_d/m_s$, etc. 
Then, by dividing the squared eq. (\ref{b-tau}) 
on eq. (\ref{sb}), we obtain: 
\be{b/s}
\frac{Y_b}{Y_s-Y_d} - (b_d + b_d^{-1}) = \frac{1}{kk'}  
\left[\frac{m_\tau}{m_\mu-m_e} - (b_e + b_e^{-1})\right] 
\ee 
to be rewritten as 
\beqn{s/b}
&&
\frac{Y_s-Y_d}{Y_b} = \frac{kk'}{Z^2} \frac{m_\mu-m_e}{m_\tau} 
\approx 0.059 \frac{kk'}{Z^2} ; 
\nonumber \\ 
&& ~~~~~~~~~
Z= \sqrt{1 - \left[ (b_e+b_e^{-1}) - kk'(b_d + b_d^{-1})\right] 
\frac{m_\mu-m_e}{m_\tau} } .
\eeqn
Substituting this expression back into eqs. (\ref{b-tau}) 
and (\ref{sb}) we get:
\be{b/tau}
\frac{Y_b}{Y_\tau} =  Z  , ~~~~~~~~~ 
\frac{Y_s-Y_d}{Y_\mu-Y_e}=\frac{kk'}{Z} ,  
\ee
Then by the RGE (\ref{RG}) we have for the physical masses:  
\beqn{bottom}
&&
m_b = \frac{R_d\eta_b}{R_e} B_t Zm_\tau = 
B_tZ\cdot (5.90 \pm 0.30) ~ {\rm GeV}, 
\nonumber \\
&& 
m_s-m_d = \frac{R_d\eta_d}{R_e} \frac{kk'}{Z} (m_\mu-m_e) 
= \frac{kk'}{4Z} \cdot (133 \pm 18)~{\rm MeV} . 
\eeqn
with uncertainties related to the value of $\al_s(M_Z)$.  

Analogously, from eqs. (\ref{dsb}) we find: 
\be{s/d}
\frac{m_s}{m_d} + \frac{m_d}{m_s} -2 = \frac{(kk')^2}{Z} 
\left(\frac{m_\mu}{m_e}+\frac{m_e}{m_\mu} -2 \right) 
\approx 204.7 \frac{(kk')^2}{Z}.  
\ee
Therefore, for $b_e,b_d$ varying within an order of magnitude, 
the value $m_s/m_d$ can vary in the `experimental' range 
$m_s/m_d =17 - 25$, if $kk'$ varies between $\frac15-\frac13$. 

Let us turn now to the quark and lepton mixing. 
We parametrize the matrices $U_d$ and $U_e$ as in the Appendix:  
$V_e=O^e_{23}O^e_{13}O^e_{12}$ and 
$U_d=O^d_{23}O^d_{13}O^d_{12}$. 
Hence, for the angles $\theta_{23}^{e,d}$ which rotate 
the left-handed states we obtain: 
\beqn{ang-23} 
&& 
\tant^{e}_{23} = 2\sqrt{b_e}\sqrt{\frac{m_\mu-m_e}{m_\tau}}
\frac{ \left[1-(b_e+b_e^{-1})\frac{m_\mu-m_e}{m_\tau}\right]^{1/2} }
{1 - 2b_e \frac{m_\mu-m_e}{m_\tau} }  ,   \nonumber \\ 
&& 
\tant^{d}_{23} = 
\frac{2}{\sqrt{b_d}}\sqrt{\frac{Y_s-Y_d}{Y_b}}
\frac{ \left[1-(b_d+b_d^{-1})\frac{Y_s-Y_d}{Y_b} \right]^{1/2} }
{1 - 2b_d^{-1} \frac{Y_s-Y_d}{Y_b} }  ,   
\eeqn 
while the right rotation angles $\theta'^{e,d}_{23}$ are 
obtained from these expressions substituting respectively 
$b_e \to b_e^{-1}$ and $b_d \to b_d^{-1}$.
For the other angles we have: 
\beqn{ang-12}
&&
\tant^{e}_{12} = 2\sqrt{a_e}\sqrt{\frac{m_e}{m_\mu}}
\frac{ \left[1-(a_e+a_e^{-1})\frac{m_e}{m_\mu} \right]^{1/2} }
{1 - 2a_e \frac{m_e}{m_\mu} }~  
\approx \frac{2}{\sqrt{c^e_{23}} }\sqrt{\frac{m_e}{m_\mu-m_e}}
\approx \frac{0.140}{\sqrt{c^e_{23}} },  \nonumber \\ 
&&
\sin\theta^e_{13} = \sqrt{a_e} \frac{s'^e_{23}}{c'^e_{23}}
\frac{\sqrt{m_em_\mu}}{m_\tau} = 
\left(\frac{m_em_\mu^2}{b_ec^e_{23} m_\tau^3}\right)^{1/2} \leq 10^{-3}, 
\eeqn 
where $a_e = c'^e_{23}/c^e_{23}$, and 
\beqn{ang-d} 
&& 
\tant^{d}_{12} = \frac{2}{\sqrt{a_d}}\sqrt{\frac{m_d}{m_s}}
\frac{ \left[1-(a_d+a_d^{-1})\frac{m_d}{m_s} \right]^{1/2} }
{1 - 2a_d^{-1}\frac{m_d}{m_s} } \approx
2\sqrt{c'^d_{23}}\sqrt{\frac{m_d}{m_s-m_d} } , \nonumber \\ 
&& 
\frac{s^d_{13}}{s^d_{23}} = \frac{1}{\sqrt{a_d}} 
\frac{s'^d_{23}}{c'^d_{23} s^d_{23}} 
\sqrt{\frac{m_d}{m_s}}\frac{Y_s}{Y_b}~,
~~~~~~~~~a_d = \frac{c^d_{23}}{c'^d_{23}}~.
\eeqn

As far as $\bY_u$ is taken diagonal, the quark mixing is 
completely determined by the form of $\bY_d$, and hence 
$V_q=U_d$. 
Thus $U_d=O^d_{23}O^d_{13}O^d_{12}$ gives the CKM matrix $V_q$ 
directly in the standard parametrization (\ref{CKM}) with 
$\delta=\pi$. Therefore, we have 
$\theta^q_{ij}=\theta^d_{ij}$ and thus: 
\beqn{V-q}
&&
|V_{cb}| = c^d_{13}s^d_{23} \approx s^d_{23},   \nonumber \\  
&&
|V_{us}|=c^d_{13}s^d_{12} \approx s^d_{12}, \nonumber \\ 
&& 
|V_{ub}|=s^d_{13} . 
\eeqn   

As for the lepton mixing, the matrix 
$V_l=V_e^\dagger = O^{eT}_{12}O^{eT}_{13}O^{eT}_{23}$ 
appears in the parametrization transposed to (\ref{CKM}). 
Hence, in standard parametrization the mixing angles read as: 
\beqn{V-l}
&&
|V_{\mu3}| = c^l_{13} s^l_{23} = 
|c^e_{12}s^e_{23} + s^e_{12}c^e_{23}s^e_{13}|\approx s^e_{23},    
\nonumber \\ 
&&
|V_{e2}|=c^l_{13} s^l_{12} = 
|s^e_{12}c^e_{23} + c^e_{12}s^e_{23}s^e_{13}|\approx 
s^e_{12} c^e_{23},    
\nonumber \\   
&&
|V_{e3}|= s^l_{13} = 
|s^e_{12}s^e_{23} + c^e_{12}c^e_{23}s^e_{13}|\approx 
s^e_{12} s^e_{23},
\eeqn   

Hence, the lepton mixing angles are expressed in terms of 
their mass ratios and asymmetry parameter $b=b_e$.\footnote{
The `critical' value of $b$ when $B_e=C$ is given by the 
equation $2b+b^{-1}=m_\tau(m_\mu-m_e)^{-1}$, and so 
$b=8.4$.
} 
In the Fig. 2 we show the $b$-dependence of  
$|V_{\mu3}|$,  $|V_{e2}|$ and  $|V_{e3}|$. 
We also show the curves for the effective oscillation 
parameters $\sin^2 2\theta_{23} = 4|V_{\mu3}|^2(1-|V_{\mu3}|^2)$ 
and $\sin^2 2\theta_{12} = 4|V_{e2}|^2(1-|V_{e2}|^2)$.  
We see that for $b=1$ the 23 mixing  angle 
($\theta^e_{23}=13.5^\circ$) is rather small for 
explaining the AN anomaly,  while the 12 mixing 
($\theta^e_{12}=4.0^\circ$) is somewhat above the upper limit 
obtained by the MSW fit of the SN data (c.f. (\ref{nu-ang})). 
However, for larger $b$, $|V_{\mu3}|$ increases roughly as 
$\sqrt{b}$ and becomes maximal around $b=8.4$, 
while $|V_{e2}|$ slowly decreases 
(roughly as $\sqrt{c^e_{23}}$). 
Thus, the AN bound, $\sin^2 2\theta_{23} >0.82$,  
requires $5<b<12$, while the MSW fit for SN data 
is recovered  at $b>7$, when $\sin^2 2\theta_{12}$ 
drops below $1.5\cdot 10^{-2}$.\footnote{ 
In fact, the value 
$\sin^2 2\theta_{\rm sol}= 1.5\times 10^{-2}$  
can be considered as acceptable 
in the context of the solar models with 
the boron neutrino flux exceeding its `standard solar model'  
value by a factor 2 or so \cite{KS}.    
}
Therefore, the relevant values of $b$ are somewhere 
between 7 and 12, in which range 
$\sin^2 2\theta_{12}$ varies from $1.5\cdot 10^{-2}$  
to $1.0\cdot 10^{-2}$, and $\sin\theta^l_{13}$ 
varies from $0.05$ to $0.08$, well below the 
upper bound of eq. (\ref{nu-ang}). 
For example, for the case $B_e=C$ we have $b=8.4$, 
and nearly maximal 23 mixing: $V_{\mu3}\approx 0.7$, 
while  $V_{e2},V_{e3}\approx 0.055$.  

Therefore, we conclude that in the basis when the neutrino 
masses are diagonal, the pattern of the lepton mixing 
required by the AN and SN anomalies can be perfectly 
described if the charged lepton mass matrix 
has a Fritzsch-like form (\ref{Fred}) with an asymmetry 
parameter $b=7-12$. We also see that in this range 
$\theta^l_{13}$ remains rather small, between $(3-5)^\circ$. 
This range, however, in the case of $\dem_{23}$ close 
to the upper bound in (\ref{AN}), can be of interest 
for the experimental search of $\nu_e\to \nu_\tau$ 
oscillation in the future CERN Neutrino Factory \cite{CERN}. 

As far as the quarks are concerned, from eqs. 
(\ref{bottom})-(\ref{s/d}) and (\ref{ang-12})-(\ref{ang-d}) 
we see that their masses and mixing angles are all 
expressed in terms of the lepton mass ratios and 
three Clebsch factors $k,k'$ and $b$. In addition, 
the physical mass of bottom depends, through the factor 
$B_t$, on the top Yukawa constant $Y_t$ at the GUT scale. 
Notice also that the expression (\ref{ang-23}) defines 
$V_{cb}$ at the GUT scale, and for obtaining its 
physical value at lower scales one has to take into 
account the factor $B_t^{-1}$, while $V_{us}$ 
and $V_{ub}/V_{cb}$ are RGE invariant.

One can see that for $k\sim k'$ and large values of $b$, 
($b=7-12$ as required from the lepton mixing)
our ansatz can give a satisfactory explanation 
also to the down quark mass and mixing pattern. 
In particular,  in Fig. 3a we show the dependence 
of the quark mixing angles on the parameter $b=b_e$ 
in the case of $b_d=b_e$, i.e. $k=k'$.\footnote{ 
In terms of the $\Phi$-dependent higher order 
operators this means that the tensor product 
$G_{23}(\Phi)\cdot \bar{H}$ and $G_{32}(\Phi)\cdot \bar{H}$ 
project the same mixture of the 5 and 45 Clebsches on 
$23$ and $32$ entries. Furthermore, 
the value $k=k'=1/2$ for the Clebsch factors 
can emerge if the effective operators 
$\frac{\Phi}{M}\bar5\bar{H}10$ is induced by the Frogatt-Nielsen 
mechanism \cite{FN} involving the heavy states in $10+\ov{10}$ 
representation, and $\Phi$ has the VEV of the form 
$\propto {\rm Diag}(0,0,0,1,1)$ \cite{ZB}. 
This means that $\Phi$ is taken as a reducible $1+24$ 
representation instead of the pure adjoint. 
Such a situation should not come up as a surprise 
if the $SU(5)$ theory is obtained from a theory with 
higher-symmetry group, say from 
$SO(10)$ or $SU(5)\times SU(5)$. Then the relevant combination 
contained in the representations 45 (for $SO(10)$) or 
$(5,5)$ (for $SU(5)\times SU(5)$) is indeed $1+24$.    
It is worth  remarking also, that another similar field 
$\Sigma\sim 1+24$ with a complementary missing 
VEV structure $\langle \Sigma\rangle \propto (1,1,1,0,0)$ 
can be used for the solution of the 
doublet-triplet splitting problem. 
}
The complementary Fig. 3b exhibites the $b$-dependence 
of the $s$-quark mass and $m_s/m_d$ ratio. 
The same Figure contains information on the 
value $Y_t$ needed for the experimental value of 
the bottom mass. Since for $k=k'$, 
we have $Z=[1-0.059(b+b^{-1})(1-k^2)]^{1/2}$,  
this provides a substantial correction to the $b-\tau$ 
Yukawa unification for large $b$. 
Therefore, in this situation rather small values 
$Y_t\sim 0.5-1$ are needed to obtain the 
correct physical mass of the bottom.    

We observe that indeed $k=1/2$ provides the 
most acceptable description for all these data, in 
comparison to the cases $k^2=1/3$ or $1/5$ which are 
also shown. 
Thus, a global inspection of Figs. 2 and 3 indicates 
that this case  with $b=7-12$ offers us a quite 
realistic picture for the quark and neutrino masses 
and mixing.

In Figs. 4 we show the same for different choice 
of the Clebsches: $k'=1$ ($b_d=k^{-1}b$) 
and $k=1/3,1/4$ and $1/5$. We see that the proper 
picture of the quark masses and mixing now requires 
rather small values of $b$, $b=3-4$, which is incompatible 
with the $b=7-12$ as it is dictated by the neutrino mixing 
pattern. One can also show that the case with $k=1$ 
($b_d=k'b$) leads to even less realistic picture.

The following remark is in order. 
Much on the same footing, one could consider 
also the ansatz when the matrices $\bG_u$ and $\bG_\nu$ 
are diagonal, $\bG_u=\bY_u^D$ and $\bG_\nu=\bY_\nu^D$, 
and the matrix $\bG$ has the form 
\be{Stech}
\bG= \rho \bY_u^D + \bA(\Phi), ~~~~~~~ 
\bA=\matr{0}{A_{12}}{0} {A_{21}}{0}{A_{23}} {0}{A_{32}}{0}  
\ee 
with $\bA$ being a matrix with the $\Phi$-field dependent 
off-diagonal (complex) entries.\footnote{
Interestingly, within the same philosophy of taking the
relevant terms diagonal between the similar multiplets 
($\bar5$'s and 10's), one could allow also the R-violating 
couplings $\la_{ik}(\Phi)\bar5_i\bar5_i 10_k$ 
which are diagonal between 5-plets for any $k$. In this case, 
the R-violating term $\la_{iik}l_id^c_iq_k$ would emerge 
in the low energy theory which  could also contribute the 
neutrino masses, while the other two terms $lle^c$ and 
$d^cd^cu^c$ would vanish by  symmetry reasons \cite{R-viol}.
} 
This pattern is  reminescent of the once popular texture 
proposed by Stech, and independently 
by Chkareuli and one of the authors \cite{Stech}.  However, 
in these papers the matrix $\bA$ has been taken antisymmetric, 
which choice was soon excluded by the experimental data. 
Committing a minimal modification of the 
original version of the ansatz \cite{Stech}, 
one could assume that the 12 block of $\bA$ remains  
antisymmetric,  $A_{12}=-A_{21}$, and $\Phi$ independent, 
whereas $\Phi$-dependent entries $A_{23}\neq A_{32}$  
are strongly asymmetric. 
The GUT models leading to this ansatz and its complete 
analysis will be presented elsewhere. It is easy to see 
that the patterns of the quark and lepton mixing 
essentially remain the same as in the Fritzsch-like ansatz 
for $\bG$ considered above  -- the presence of the diagonal 
22 and 11 entries in $\bG$ with the hierarchy stepped as 
$Y_u:Y_c:Y_t$ will lead only to small corrections.  
However, the CP-violating phase in this case would not vanish.

\section{Give Fritzsch texture a Chance 
also for $\bY_u$ and $\bY_\nu$ } 

In the above we have assumed 
that the matrix $\bG$ has a Fritzsch-like texture while 
$\bG^u$ and $\bG^\nu$ are diagonal. However, in the context 
of models in which the form of $\bG$ is fixed by some 
underlying horizontal symmetry reasons, it would 
seem more natural that $\bG^u$ and $\bG^\nu$ also have a 
similar form: 
\be{Fr-nu}
\bG_{u}= \matr{0}{G^u_{12}}{0} {G^u_{21}}{0}{G^u_{23}}
{0}{G^u_{32}}{G^u_{33}},
~~~~~~
\bG_{\nu}=\,\matr{0}{G^\nu_{12}}{0}
{G^\nu_{12}}{0}{G^\nu_{23}}
{0}{G^\nu_{23}}{G^\nu_{33}},
\ee
Clearly, if the entries in $\bG_u$ are  $SU(5)$ singlets, 
this matrix should be symmetric. More generally, for 
$\Phi$ dependent entries, this is not true anymore, 
since the symmetric contribution can be induced 
by terms containing the tensor product 
$\Phi\cdot H$ in the 5-channel and 
the antisymmetric ones by terms containing  $\Phi\cdot H$ 
in the 45-channel.  As for $\bG_\nu$, 
clearly only the symmetric terms are relevant for 
the neutrino mass matrices. 
For simplicity, in the next we 
assume that also $\bG_u$ is symmetric.

Such a scenario would provide some different features. 
In the model with diagonal $\bG^u$ and $\bG^\nu$ we 
have been forced to take too big an asymmetry between 
the 23 and 32 entries of $\bG$, $b_{e,d}\lsim 10$. 
In the case in which also  $\bG^u$ and $\bG^\nu$  have the 
form (\ref{Fr-nu}), smaller values of $b_{e,d}$ can suffice 
since now the mixing angles will be contributed also 
by the unitary matrices $U_u$ and $U_\nu$: 
$V_q=U_u^\dagger U_d$ and $V_l=U_e^\dagger U_\nu$. 
Therefore, 
for the CKM mixing angles we have: 
\be{Vcb}
|V_{cb}|=s^q_{23}\approx  
\left|s^d_{23}-e^{i\varphi}s^u_{23}\right|, ~~~~~~~~
|V_{us}|=s^q_{12} \approx
\left|s^d_{12}-e^{i\delta}s^u_{12}\right|, ~~~~~~~~ 
\left|\frac{V_{ub}}{V_{cb}}\right|\approx s^u_{12} 
\ee
where the phases $\varphi$, $\delta$ etc are combinations 
of the independent phases in the Yukawa matrices. 
The expressions for $\theta^d_{23}$ 
and $\theta^d_{12}$ are the same as in  
(\ref{ang-23}) and (\ref{ang-d}), 
while $\theta^u_{23}$, $\theta^u_{12}$ are the analogous angles 
diagonalizing $\bY_u$:  
$\tan\theta^u_{23}=\sqrt{Y_c/Y_t}$ and 
$\tan\theta^u_{12}=\sqrt{m_u/m_c}$. 
The dependence of $\theta^d_{23}$ and $\theta^d_{12}$
on the parameter $b$ for the cases $k=k'$ and $k'=1$ 
can be read out respectively from Figs. 3a and 4a, 
and for small values $b$ the contribution of $s^d_{13}$ 
in $V_{ub}/V_{cb}$ can be neglected. 

Therefore, by varying  the phase $\varphi$  from $0$ to $\pi$, 
the value of the 23 mixing angle in the CKM matrix can 
vary between its minimal and maximal possible values:  
\be{pm}
\theta^{q(\mp)}_{23} = \theta^d_{23} \mp \theta^u_{23} 
\ee
Analogously, for the leptonic mixing we have 
\be{pm-nu23}
\theta^{l(\mp)}_{23} = \theta^e_{23} \mp \theta^\nu_{23} 
\ee
where $\tan\theta^\nu_{23}= \sqrt{m_2/m_3}$. Thus, 
for the range of the neutrino masses indicated in 
(\ref{nu-spectr}) we obtain 
$\theta_{23}^\nu= (12.2_{-3.7}^{+8.6})^\circ$.
In case of  moderate asymmetry in $\bY_{d,e}$, 
the entries in (\ref{pm}) are big as compared to the 
experimental value of $\theta^{q}_{23}$ while each 
of the entries in (\ref{pm-nu23}) is too small for the 
magnitude of $\theta^{l}_{23}$ required by the AN oscillation. 
However, by properly tuning the phases, 
 $\theta^{q}_{23}$  can get close to 
$\theta^{q(-)}_{23}=\theta^d_{23} - \theta^u_{23} $ 
while $\theta^{l}_{23}$ can approach 
$\theta^{l(+)}_{23}=\theta^e_{23} + \theta^\nu_{23} $,  
In other words, the angles in the quark and 
lepton sectors could have 
negative and positive interference, respectively.    
Therefore, even for smaller values of $b_{e,d}$, one could 
achieve a proper fit of the mixing angles. 

Nevertheless, it is well known that 
the case $b_{e,d}=1$, corresponding to  the original 
Fritzsch texture, is fully excluded. 
In this case we have 
$s^d_{23}\approx \sqrt{Y_s/Y_b}$ which implies a  
sharp conflict between the values of $|V_{cb}|$ and the large top mass 
$M_t$.  
Namely, the Fig. 5 shows that for the most conservative 
bound $M_t>160$ GeV, even the least possible value 
$\theta^{q(-)}_{23}=\theta^d_{23} - \theta^u_{23} $ 
exceeds its experimental range by a factor 2 or so 
(see the dotted curves).

Neither in the leptonic sector it is possible to fully reproduce 
 the desired pattern (\ref{nu-ang}). 
For $b_e=1$ we have $\tan\theta^e_{23} = \sqrt{m_\mu/m_\tau}$, 
i.e. $\theta^{e}_{23}=13.6^\circ$. 
Therefore, only the maximal value 
$\theta^{l(+)}_{23}=\theta^e_{23} + \theta^\nu_{23} $ 
can marginally satisfy the lower bound of 
(\ref{nu-ang}).\footnote{This possibility was 
remarked in ref. \cite{Pati}. }
On the other hand, now the 12 mixing is contributed also 
by the angle $\tan\theta^\nu_{12}=\sqrt{m_1/m_2}$ and 
therefore the unknown value of $m_1$ makes invalid 
the expression for $s_{12}^l$.

However, by taking the matrices $\bY_{e,d}$ somewhat asymmetric 
in the 23 block, the situation can improve significantly. 
Already the choice $b_{e,d}=2$ can suffice (such a 
possibility was suggested e.g. in ref. \cite{ICTP}, where 
this factor 2 was obtained as a horizontal $U(3)_H$ symmetry 
breaking Clebsch). In this case we have 
$s^d_{23}\approx \sqrt{(Y_s/2Y_b)}$ and then 
the value of 
$\theta^{q(-)}_{23}=\theta^d_{23} - \theta^u_{23} $ 
can perfectly  agree with the large $M_t$, as it is demonstrated 
in the Fig. 5 (see the dashed curves). 
On the other hand, now we obtain also 
$s^e_{23}\approx \sqrt{(2m_\mu/m_\tau)}$,  i.e. 
$\theta^{e}_{23}=23^\circ$, and thus 
the value of 
$\theta^{l(+)}_{23}=\theta^e_{23} + \theta^\nu_{23} $ 
can fit well into the range (\ref{nu-ang}) required by 
AN oscillation.  

In the above we have assumed that $k=k'$, 
where $k$ and $k'$ are the $SU(5)$ breaking Clebsches 
defined as in eq. (\ref{Clebsch}). However, in general 
these Clebsch factors have no reason to be the same.  
Nevertheless, for the dominant contributions $\theta^d_{23}$ 
and $\theta^e_{23}$ to the 23 quark and leptonic  mixing angles  
(\ref{pm}) and (\ref{pm-nu23}) 
it emerges the  "seesaw" like product rule: 
\be{rule} 
\tan\theta_{23}^d \tan\theta_{23}^e \simeq 
\left(\frac{k}{k'}\right)^{1/2} 
\left(\frac{m_\mu m_s}{m_\tau m_b}\right)^{1/2}. 
\ee
Barring unusual conspiracies, 
one can expect both $k$ and $k'$ to be 
$\leq 1$ and then,  bearing also in mind 
that $kk'\simeq (Y_s/Y_\mu)\sim 1/3-1/4$,  
the factor $(k/k')^{1/2}$ appears to be $\sim 1$. 
Even in rather extreme case $k=1/3$ and $k'=1$, 
this factor is of about 0.6.

At this point some natural questions emerge, concerning the 
reasons for 
fermion Yukawa constants to have such Fritzsch-like textures, 
and the  motivations underlying  our assumptions. 
However, it is already known in the literature that such 
textures can naturally occur due to a horizontal symmetry 
(e.g. \cite{PLB85,ICTP}). Some very predictive schemes,
based on the concept of the horizontal $U(3)_H$ symmetry 
\cite{su3J}, and 
fully demonstrating the features discussed here, 
 will be presented elsewhere. Here we confine 
ourselves to the  discussion of  theories with a simpler 
field content based on the horizontal $U(2)_H$ symmetry 
\cite{BDH,BH}.\footnote{
Applications of the $U(2)$ horizontal symmetry for 
the neutrino mass and mixing pattern were discussed also 
in refs. \cite{U2-nu}, but in a different context. 
} 
 
Indeed, consider a theory based on the $SU(5)\times U(2)_H$ 
symmetry where $U(2)_H=SU(2)_H\times U(1)_H$ 
stands for a horizontal symmetry unifying the 
first two families in a doublet: $\bar5_\al=(\bar5,2)$, 
$10_\al = (10,2)$, $\al=1,2$, with an $U(1)_H$ hypercharge $H=-1$, 
while the third family ($\bar5_3,10_3$) is a $U(2)$ singlet 
with $H=0$ \cite{BH}. The Higgs 5-plets are singlet of 
$U(2)_H$, $H\sim (5,1)$ and $\bar{H}\sim (\bar5,1)$, and 
theory contains also some amount of $SU(5)$ singlets $S\sim (1,1)$ 
and adjoints $\Phi \sim (24,1)$, with VEVs of order of 
$M\sim M_G\simeq 10^{16}$ GeV. 
For the horizontal symmetry breaking 
one can introduce the simplest set of the Higgs superfields:  
a doublet $\phi^\al=(1,2)$ with $H=1$ and 
a singlet $A=(1,1)$ with $H=2$, having  nonzero VEVs 
$\langle \phi^\al \rangle=V_\phi \delta^\al_2 $ and 
$\langle A \rangle= V_A$, smaller than $M_G$. 

The third generation can get masses through the 
Yukawa couplings 
$H10_310_3 + \bar{H}10_3\bar5_3$ 
while the masses of the first two generations emerge 
from the effective operators induced by the 
Frogatt-Nielsen mechanism after integrating out 
some heavy vector-like states in representations 
$10+\ov{10}$ and $\bar5+5$ \cite{FN}.  
Following ref. \cite{BH}, the latter can be chosen 
as the $U(2)_H$ doublets $T^\al +\bar{T}_\al$ (Ten-plets) 
and $\bar{F}^\al + F_\al$ (Five-plets). For the neutrino 
masses, we add also the $SU(5)$ singlet fermions 
(right-handed neutrinos) $N^\al+\bar{N}_\al$, $a=1,2$, 
and $N+\bar{N}$. All these get large ($M\sim M_G$) masses 
from couplings to the fields $S$ and $\Phi$: 
\be{FN-heavy}
(S+\Phi)T^\al \bar{T}_\al + (S+\Phi)\bar{F}^\al F_\al + 
S (N^\al\bar{N}_\al + N^2 + N\bar{N} + \bar{N}^2) 
\ee
Therefore, the heavy masses in the Ten- and Five-plets 
exhibite the $SU(5)$ breaking between various 
$SU(3)\times SU(2)\times U(1)$ fragments with different 
$U(1)$ hypercherges $Y$ -- they are 
respectively $M_T(1+ x_T Y)$ and $M_F(1+ x_F Y)$.  
Then the mass terms of the first two generations emerge 
by the {\it seesaw} mixing with the heavy ones: 
\beqn{FN-mix} 
&&
H10_\al T^\al + \bar{H}\bar5_\al T^\al 
+ \bar{H} 10_\al\bar{F}^\al 
+ H\bar5_\al N^\al + H\bar5_3 N,  \nonumber \\ 
&& 
A\varepsilon^{\al\beta}(\bar{T}_\al 10_\beta + F_\al\bar5_\beta) 
+ \phi^\al (\bar{T}_\al10_3  + F_\al\bar5_3) 
+ \phi^\al\bar{N}_\al(N+\bar{N}) 
\eeqn
For example, the effective operator for the neutrino masses 
appears in a form: 
\be{op-nu}
\frac{HH}{S}\bar5_3\bar5_3 + 
\frac{HH\phi^\al}{S^2}\bar5_\al\bar5_3 .  
\ee   
Therefore, the mass of the heaviest neutrino $\nu_3$ 
could naturally emerge in the needed range around 
$m_3\sim 4\cdot 10^{-2}$ eV (c.f. (\ref{nu-spectr})). 
In fact, literally in the context of the effective 
operator (\ref{op-nu}) this implies 
$\langle S\rangle \sim 10^{15}$ GeV rather than the 
GUT scale $M_G\simeq 10^{16}$ GeV, but in terms of 
the renormalizable couplings (\ref{FN-heavy}) this mismatch 
can be easily understood e.g. due to some difference 
of the relevant coupling constants.\footnote{Notice, 
that all couplings contained in the theory are trilinear, 
and thus they invariant under the continuous $R$-symmetry 
with all superfields having $R$-charge 1 and the superpotential 
having the $R$-charge 3. From one side, this forbides the 
direct mass terms of superfields $T$, $F$ and $N$ to be 
of the order of the Planck scale. 
On the other hand, the superpotential 
terms of the GUT superfields $S$ and $\Phi$ should 
contain only the trilinear terms $S^3+S\Phi^2+\Phi^3$. 
Then the mass parameter giving rise to their VEVs 
could emerge from the coupling $SQ\bar{Q}$ to the 
fermions $Q$, $\bar Q$ of the strongly coupled gauge sector,  
through the linear term $\Lambda^2S$ 
emerging due to the dynamical condensation  
$\langle \bar{Q}Q\rangle =\Lambda^2$. 
}  

Hence, after integrating out the heavy states, we arrive 
to the following Fritzsch-like textures for the 
the Yukawa couplings: 
\be{Fr-U2ed}
\bY_d = \matr{0}{-A_d}{0} {A_d}{0}{B'_{d}} {0}{B_{d}}{C}, 
~~~~~~ 
\bY_{u}= \matr{0}{-A_u}{0} {A_u}{0}{B'_u}{0}{B_u}{C_u},
\ee
and 
\be{Fr-U2}
\bY_e = \matr{0}{A_e}{0} {-A_e}{0}{B'_{e}} {0}{B_{e}}{C}, 
~~~~~~ 
\bY_{\nu}= \matr{0}{0}{0}{0}{0}{B_\nu}{0}{B_\nu}{C_\nu},
\ee 
with 33 elements $C_u$ and $C_\nu$ naturally being $\sim 1$, 
while $C$ should be somewhat smaller unless $\tan\beta$ is 
very large.  
Interestingly, when $r=M_T/M_F\ll 1$ and $x_T\ll 1$, 
the other Yukawa entries have the following hierarchy: 
\beqn{entries}
&&
A_d\approx A_e=A\sim \frac{V_A}{M_T}, ~~~~ A_u\sim x_T A, 
\nonumber \\
&&
B'_u\approx B_u \sim \frac{V_\phi}{M_T}, ~~~~
B'_d\approx B_e \sim \frac{V_\phi}{M_T},   
\nonumber \\
&&
B_\nu \sim V_\phi/M_N, ~~~~~
B_d, B'_e \sim \frac{V_\phi}{M_F}\sim  r B_e. 
\eeqn
Thus, we have $k'\approx 1$ and $b\sim 1/r$, since  the 
asymmetry in the 23 block of $\bY_{e,d}$ is due to 
the mass difference between heavy Ten- and Five-plets. 
In addition, if $x_F\sim 1$, then $k=B_d/B'_e\neq 1$ 
and in particular it could be chosen around $1/3-1/4$. 
The implications of such ansatz for the quark masses and 
mixing was discussed in great details in ref. \cite{BH}. 

We would like to add the following remark. 
As was already said, now one 
can naturally achieve the large 23 mixing in the lepton 
sector even in the case of smaller asymmetry 
in $\bY_e$,  already starting from $b\sim 1$, 
since now besides the ($b$-dependent) charged 
lepton angle $\theta^e_{23}$ it is contributed also 
by the neutrino angle $\theta^\nu_{23}$
from the matrix $\bY_\nu$ 
($\tan\theta^\nu_{23} = \sqrt{m_2/m_3}$). 
In addition, in this model we have $A_\nu=0$, 
since there can be no antisymmetric entry for the neutrino 
Majorana masses. Hence, the eigenstate 
$\nu_1$ remains massless, $m_1=0$, and therefore the 12 
mixing angle in leptonic sector can be predicted. 

In particular, now we have 
$|V_{e2}|=s^l_{12}\approx s^e_{12}c^l_{23}$, 
and assuming the positive interference in 
eq. (\ref{pm-nu23}), we find that $\theta^l_{23}$ falls 
in the range needed for AN oscillation,  
$\theta^l_{23}=(33-57)^\circ$,   
already for moderate values of $b$ for which 
$c^e_{23}\approx 1$ and so $s^e_{12}\approx \sqrt{m_e/m_\mu}$.  
This implies that for the angle $\theta^l_{23}$ 
in the range needed for the AN oscillation 
(i.e. $c^l_{23}=0.83-0.55$), 
the angle $\theta^l_{12}$ is entirely contained 
in a range required by SN: 
$\sin^2 2\theta^l_{12} \approx 4(m_e/m_\mu) (c^l_{23})^2 = 
(1.4- 0.6)\times 10^{-2}$, and moreover 
it can approach also the best MSW fit value of 
$\sin^2 2\theta_{\rm sol}$  \cite{BKS}. 
The $b$-dependence of the lepton 
mixing pattern for the interval 
$\theta^\nu_{23}=(12.2^{+8.6}_{-3.7})^\circ$ which 
covers uncertainties in the neutrino masses  
(\ref{nu-spectr}) is shown in Fig. 6.

\section{Discussion and outlook}

We have argued that the neutrino mixing pattern required by 
the solutions of the AN and SN anomalies can be obtained 
in a rather natural way in the context of the $SU(5)$ grand 
unification, assuming that contributions to the leptonic mixing 
angles emerge completely or domininantly 
from the charged-lepton mass matrices which 
have Fritzsch-like textures with strongly asymmetric 23 block 
and nearly symmetric 12 block.  
In this case, as far as $\bY_d\sim \bY_e^T$ modulo  
$SU(5)$ symmetry breaking Clebsch factors,  
the large value of the 23 leptonic mixing angle  
can be nicely linked to the small value of the 23 mixing angle 
in the quark sector. 
Namely, for the dominant contributions 
to the quark and lepton angles of 23 mixing  
it emerges the  "seesaw" balance rule (\ref{rule23}). 
And vice versa, the small 12 leptonic mixing 
can have a natural link to the significant 
12 mixing in quarks, expressed as in eq. (\ref{rule12}). 

One has to remark, however, that besides the AN and SN anomalies, 
there is another, though rather controversial, indication 
for the neutrino oscillations: the LSND anomaly \cite{LSND}. 
Namely, the data collected by the LSND collaboration 
indicate neutrino oscillations in both the  
$\bar{\nu}_\mu-\bar{\nu}_e$ and 
$\nu_\mu-\nu_e$ channels. Although these are almost in  
conflict with the data of the KARMEN experiment \cite{KARMEN}, 
which excludes the bulk of the relevant parameter region, 
the following range is still allowed:
$\delta m^2_{e\mu}= (0.2-0.6)~\mbox{eV}^2$ ,
~~
$\sin^2 2\theta_{e\mu}= (0.4-4)\cdot 10^{-2}$.

If the LSND anomaly will be indeed  confirmed in 
future experiments, then three standard
neutrinos  $\nu_e,\nu_{\mu}$ and $\nu_{\tau}$ would not suffice
for reconciling AN and SN solutions to the parameter range 
required by the LSND. 
Since the existence of the fourth active neutrino
is excluded by the LEP measurements of the invisible decay width of
$Z$-boson, one has to introduce an extra light sterile
neutrino $\nu_s$ \cite{calmoh}. 
Then several exotic textures \cite{SV} can be considered 
for accomodating all the data and one 
has also to think of the physical reasons for the existence 
of the light sterile neutrinos. For some older and recent works 
on these directions see e.g. refs. \cite{mirror}.

\vspace{6mm}
{\Large \bf Acknowledgements} 
\vspace{3mm}

We would like to thank Alexei Smirnov and Francesco 
Vissani for useful discussions, and Andrea Brignole  
for valid observations.  
Z.B. thanks the high energy theory groups of the 
ICTP and SISSA for the hospitality in Trieste 
during the period when this work has been done. 

\vspace{5mm} 
{\it Note added at replacement:} a week after our paper 
has been submitted to the hep-ph E-print Archive, there 
appeared also a work by K. Hagiwara and N. Okamura 
(hep-ph/9811495) with similar considerations. 

\vspace{10mm}

{\Large \bf  Appendix}
\vspace{5mm}

Consider the Yukawa couplings of any type of fermions 
$f^c_i \bY^{ij} f_j \phi$, $f=u,d,e$, 
with the matrix $\bY$ having the Fritzsch-like form with 
generically complex elements: 
\be{Ap-1}
\bY = \matr{0}{A'e^{i\al'}}{0} {Ae^{i\al}}{0}{B'e^{i\beta'}} 
{0}{Be^{i\beta}}{C} 
\ee
obeying the hierarchy $C > B,B'> A,A'$  
(without loss of generality, the 33 element can be taken real). 
It can be brought to the diagonal form by bi-unitary transformation:  
\be{Ap-2}
U'^T \bY U  = \bY_D = 
\matr{Y_1}{0}{0} {0}{Y_2}{0} {0}{0}{Y_3} 
\ee
where $Y_3 \gg Y_2 \gg Y_1$ are the Yukawa eigenvalues for 
the physical fermions of three families. 
The unitary matrices can be parametrized as 
$U = PO$ and $U' = P'O'$, 
where the phase transformations 
\be{Ap-3}
P' =\matr{e^{i(\pi-\al'+\beta)}}{0}{0} {0}{e^{-i\beta'}}{0} 
{0}{0}{1}, 
~~~~~ 
P =\matr{e^{i(\pi-\al+\beta')}}{0}{0} {0}{e^{-i\beta}}{0} 
{0}{0}{1}, 
\ee 
bring $\bY$ to the real form 
\be{Ap-3a} 
P'\bY P = \tilde{\bY} = \matr{0}{-A'}{0} {-A}{0}{B'} {0}{B}{C} , 
\ee 
which further can be diagonalized by the bi-orthogonal 
transformation $O'^T\tilde{\bY} O$, with the matrix 
$O$ rotating the left states $f$ parametrized as 
\be{Ap-3b}
O = O_{23} O_{13} O_{12} \equiv 
\matr{1}{0}{0} {0}{c_{23}}{s_{23}} {0}{-s_{23}}{c_{23}} 
\matr{c_{13}}{0}{-s_{13}} {0}{1}{0} {s_{13}}{0}{c_{13}}  
\matr{c_{12}}{s_{12}}{0} {-s_{12}}{c_{12}}{0} {0}{0}{1} , 
\ee 
and analogously $O'=O'_{23}O'_{13}O'_{12}$ 
for the right states $f^c$. 
Notice, that (\ref{Ap-3b}) gives the matrix $O$ 
immediately in the standard parametrization (\ref{CKM}) 
with $\delta=\pi$. 

Let us compute now these rotation angles, using the 
fact that $O^T \tilde{\bY}^T\tilde{\bY} O = 
O'^T\tilde{\bY} \tilde{\bY}^TO' = \bY_D^2$.  
For the sake of accuracy, below we maintain 
the corrections of the order $\eps\sim Y_1/Y_2$, $Y_2/Y_3$, but 
neglect the ${\cal O}(\eps^2)$ ones 
($\sim Y_1/Y_3$, $(Y_2/Y_3)^2$ etc.).  
As for the elements in (\ref{Ap-3a}), in first approximation 
we estimate that 
\be{estimates}
C\sim Y_3, ~~~~~ BB'\sim Y_2Y_3, ~~~~~ AA'\sim Y_1Y_2 
\ee 
so that $C^2 : BB' : AA' \sim 1 : \eps : \eps^3$. 
We assume that $B$ and $B'$ can be substantially different, 
i.e. asymmetry parameter $b=B/B'$ can be large.  
In particular, $B$ can be $\sim C$, in which case $B'\sim \eps C$ 
and thus $b\sim 1/\eps$. As for $A$ and $A'$, we assume 
that they have no big asymmetry, and thus 
$A\sim A' \sim \eps^{3/2}C$. 

We start with the 23 rotation to diagonalize  
the lower 23 block of $\tilde{\bY}$:
\be{rot23} 
O'^T_{23}\bY^{(0)} O_{23} = \bY^{(1)} = 
\matr{0}{-c_{23}A'}{-s_{23}A'}{-c'_{23}A}{-y_2}{0}
{-s'_{23}A}{0}{y_3} .
\ee
Then we have 
\be{rot23-1} 
BB'=\frac{1}{b}B^2=y_2y_3 , ~~~~~ 
C^2 = y^2_3 + y^2_2 - (B^2 + B'^2) = 
y^2_3\left[1-(b+b^{-1})\frac{y_2}{y_3} + 
\frac{y^2_2}{y^2_3} \right]
\ee 
where $b=B/B'$, and for  the left ($f$) and right 
($f'$) rotation angles we get respectively: 
\beqn{rot23-angles} 
&&
\tant_{23} = \frac{2BC}{C^2 - B^2 + B'^2} = 
2\sqrt{\frac{by_2}{y_3}} 
\frac{ \left[1-(b+b^{-1})\frac{y_2}{y_3}\right]^{1/2}  } 
{1 - 2b \frac{y_2}{y_3} }  , \nonumber \\ 
&&
\tant'_{23} = \frac{2B'C}{C^2 + B^2 - B'^2} = 
2\sqrt{\frac{y_2}{by_3}} 
\frac{ \left[1-(b+b^{-1})\frac{y_2}{y_3}\right]^{1/2}  } 
{1 - 2b^{-1} \frac{y_2}{y_3} }  
\eeqn 
Thus, the expressions for $\theta_{23}$ and $\theta'_{23}$ are 
obtained from each other by changing $b \to b^{-1}$. 
The diagonal elements $y_{2,3}$, up to small corrections 
detected below,  
coincide with the Yukawa eigenvalues $Y_{2,3}$ in (\ref{Ap-2}), 
$y_{2,3}\simeq Y_{2,3}$.  Therefore, with increasing $b$ from 
1 to $\sim\eps^{-1}$, 
$s_{23}$ increases from $\sqrt{\eps}$ to 1, 
while at the same $s'_{23}$ decreases from $\sqrt{\eps}$ to  
$\sqrt{\eps/b} \sim \eps$, so that we always have 
$s_{23}s'_{23} \sim \eps$. 


As a next step, we rotate out the 13 block of $\bY^{(1)}$:  
\be{rot13} 
O'^T_{13}\bY^{(1)} O_{13} = \bY^{(2)} = 
\matr{-\Delta y_1}{-c'_{13}c_{23}A'}{0} 
{-c_{13}c'_{23}A}{-y_2}{s_{13}c'_{23}A}
{0}{s'_{13}c_{23}A'}{y_3 + \Delta y_3} . 
\ee
where the angles $\theta_{13}$ and $\theta'_{13}$ are small: 
\be{mix13}
s_{13} = s'_{23}\frac{A}{y_3}, ~~~~~ 
s'_{13} = s_{23}\frac{A'}{y_3}
\ee
so that $s_{13}s'_{13}\sim \eps^4$, and 
\be{Delta}
\Delta y_1 = s_{23}s'_{23} \frac{AA'}{y_3} 
\sim \eps^2 Y_1, ~~~~ 
\Delta y_3 = \frac{1}{2y_3}(s_{23}^2 A'^2 + s'^2_{23}A^2) 
< \eps^3 Y_3
\ee
Therefore, with a great precission, $c_{13}=c'_{13}=1$, 
and $\Delta y_{1,3}$ are negligible. 

Finally, we bring the 12 block of $\bY^{(2)}$ 
to the diagonal form, which step practically accomplishes 
the diagonalization procedure, since 
$\bY^{(3)} = {\rm Diag}(1,-1,1)\times \bY_D$ 
with a very good precission:  
\be{rot12} 
O'^T_{12}\bY^{(2)} O_{12} = \bY^{(3)} = 
\matr{y_1}{0}{s_{12}s_{13}c'_{23}A}
{0}{-y_2 + \Delta y_2}{c_{12}s_{13}c'_{23}A}  
{s'_{12}s'_{13}c_{23}A'}{c'_{12}s'_{13}c_{23}A'}{y_3} 
\approx \matr{Y_1}{0}{0} {0}{-Y_2}{0} {0}{0}{Y_3}. 
\ee
Therefore, up to corrections ${\cal O}(\eps^2)$, 
we have $y_3=Y_3$,  $y_2-\Delta y_2 =Y_2$ and $y_1=Y_1$. 
The next series of subsequent rotations would bring at most 
${\cal O}(\eps^2)$ corrections to the mixing angles and 
Yukawa eigenvalues. 
Furthemore, we obtain the following relations:
\beqn{rot12-1} 
&&
c_{23}c'_{23}AA'= \frac1a (c'_{23}A)^2 = Y_1Y_2 , 
\nonumber \\ 
&&
y^2_2 = Y_2^2 + Y_1^2 - (c^2_{23}A'^2 + c'^2_{23}A^2) =  
Y_2^2\left[1-(a+a^{-1})\frac{Y_1}{Y_2} + 
\frac{Y^2_1}{Y_2^2} \right]
\eeqn 
where $a=(A c'_{23}/A'c_{23})$.  
Now $Y_2$ and $Y_1$ practically coincide with the 
light generation of the Yukawa constants: e.g. 
$Y_{1,2}=Y_{e,\mu}$ for the charged leptons. 
Thus we have 
\be{Y2}
y_2 = Y_2 - \frac12(a+a^{-1})Y_1 = Y_2-Y_1[1+ \frac12(1-a)^2 +...]
\ee 
since for $a\approx 1$, we have 
$a+a^{-1} = 2 + (a-1)^2+...$, 
and hence $y_2\approx Y_2-Y_1$.  
In conclusion th expressions for the mixing angles are the following:
\beqn{rot-angles} 
&&
\tant_{23} = 2\sqrt{b}\sqrt{\frac{Y_2-Y_1}{Y_3}} 
\frac{ \left[1-(b+b^{-1})\frac{Y_2-Y_1}{Y_3}\right]^{1/2} }  
{1 - 2b \frac{Y_2-Y_1}{Y_3} }  , \nonumber \\ 
&&
\tant_{12} = 
2\sqrt{a}\sqrt{\frac{Y_1}{Y_2}} 
\frac{ \left[1-(a+a^{-1})\frac{Y_1}{Y_2} \right]^{1/2} } 
{1 - 2a \frac{Y_1}{Y_2} }  ,  \nonumber \\
&& 
\sin\theta_{13} = s'_{23}\frac{A}{Y_3} = 
\sqrt{a}\frac{s'_{23}}{c'_{23}}\frac{\sqrt{Y_1Y_2}}{Y_3} 
\eeqn 
while the expressions for $\theta'_{23}$, $\theta'_{12}$ 
and $\theta'_{13}$ are obtained from the above ones 
by changing $b\to 1/b$, $a\to 1/a$ 
and $s'_{23} \to s_{23}$ ($c'_{23} \to c_{23}$).

\newpage

{\bf Figure Captions.}
\vspace{0.5cm}

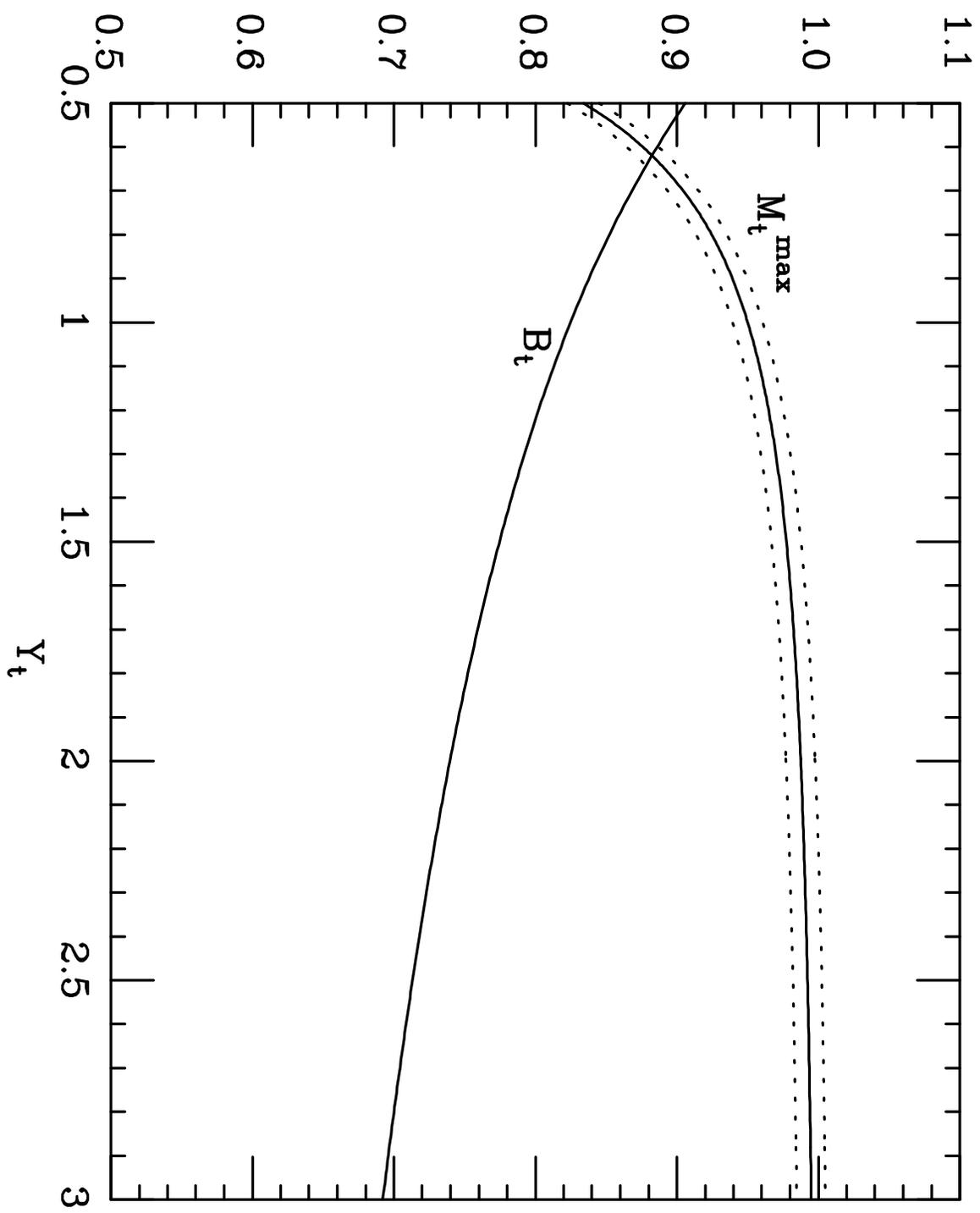
Fig. 1. $~$
The decreasing  solid line is the scaling factor $B_t$ 
as a function of $Y_t$ for $\alpha_s =0.118$ 
(the sensitivity of $B_t$ with respect to 
uncertainties in $\alpha_s(M_Z)$ is very low, around per cent 
or so).
We also show  the $Y_t$ dependence of the top pole mass 
$M^{max}_t$ (at $\sin\beta=1$)
for $\alpha_s =0.118\pm 0.005$. 

\vspace{0.4cm}

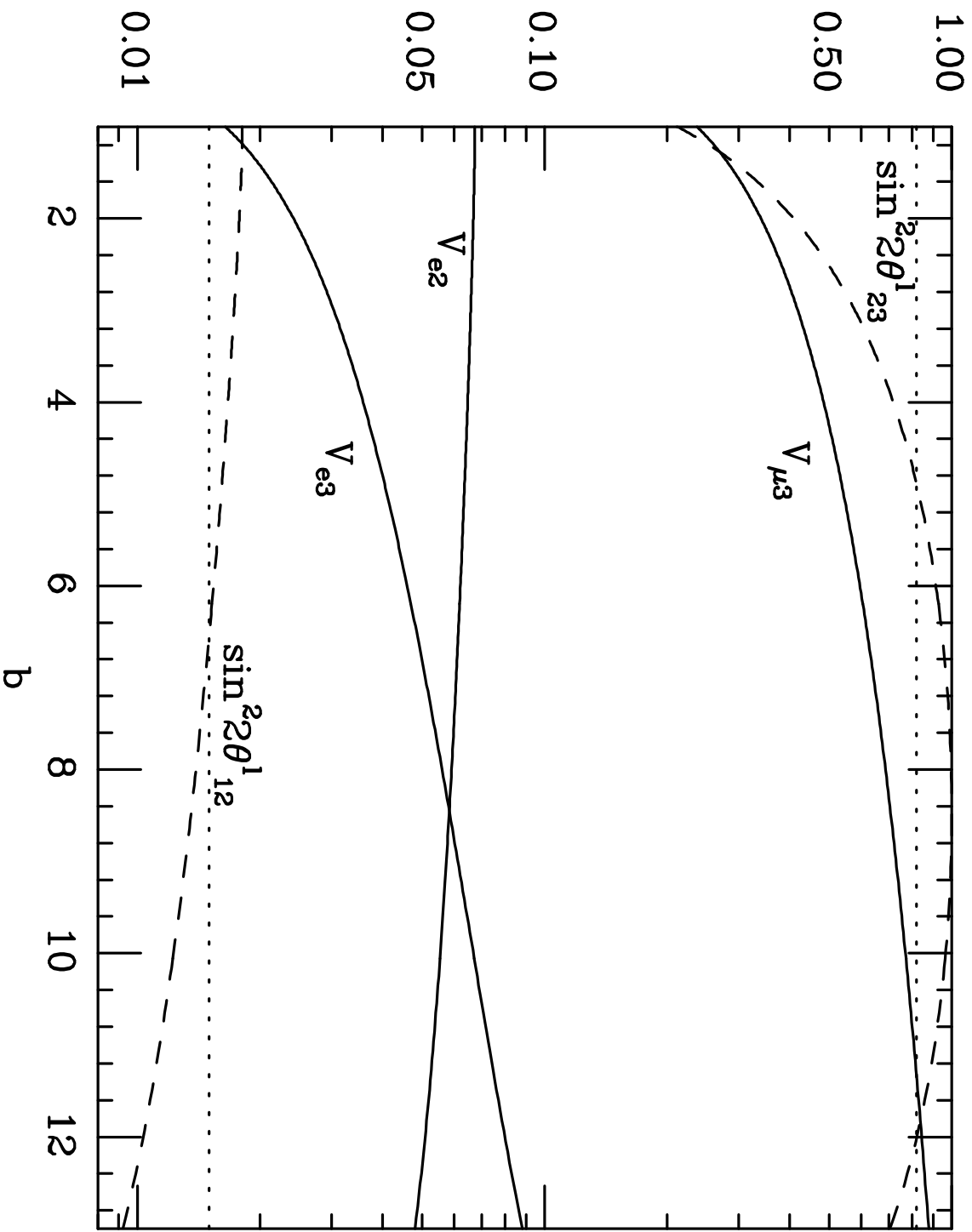
Fig. 2. $~$ The lepton mixing angles 
$V_{e2}, V_{e3}$ and $V_{\mu 3}$    
as functions of $b$ (solid). 
The oscillation parameters 
$\sin^22\theta^l_{23}$ and $\sin^22\theta^l_{12}$  
are also shown (dash), and the experimental 
limits $\sin^22\theta^l_{23} >0.82$ and 
$\sin^22\theta^l_{12}< 1.5 \cdot10^{-2}$ 
are delimited by the dotted lines.

\vspace{0.4cm}


Fig. 3. $~$ The quark mixing angles (Fig 3a) 
and the down-quark masses (Fig. 3b) as functions of $b=b_e$ 
in the case $b_d=b$ (i.e. $k'=k$). 
Three different values are considered:  $k^2=1/4$ (solid),
$k^2=1/3$ (dott) and $k^2=1/5$ (dash).
All mixing angles shown in Fig. 3a are evaluated at 
the GUT scale.
In Fig. 3b.  the strange mass 
$m_s(1~{\rm GeV})$ is shown in units of 100 MeV and the
 ratio $m_s/m_d$ in units of 20. 
In Fig. 3b we also show iso-contours for $m_b=4.25$ GeV 
in the $b-Y_t$ plane (for $\alpha_s=0.118$). 
In fact, these curves signify  
the implicit dependence of $Y_t$ on $b$ in the context 
of the ansatz. The corresponding values of $B_t^{-1}$ 
needed to rescale $s^d_{23}$ are to be taken from the Fig. 1. 
The effect of the uncertainties in $\alpha_s$ and  
$m_b$ for $Y_t$ is presented in Fig. 3c  
(for  $k=k'=1/2$). 
The solid iso-contours correspond to $m_b=4.25\pm0.15$ 
for $\alpha_s=0.118$, and marginal cases 
$m_b=4.1$, $\alpha_s=0.123$, and $m_b=4.4$, $\alpha_s=0.113$ 
are outlined by the dotted contours.

\vspace{0.4cm}


Fig. 4. $~$ 
The same as in the Fig. 3a and 
Fig. 3b, for $k'=1$ and $k=1/4$ (solid) $k= 1/3$ (dott) 
and  $k=1/5$ (dash). 
Also here $b=b_e$  (and $b_d= \frac{1}{k}b$). 
In this case the factor $Z$ is close to 1 and thus 
$Y_b=Y_\tau$, which requires larger values of $Y_t$, 
at the margin of the perturbative regime.
This is in drastic contrast to what is 
exhibited in Fig. 3b (case $k=k'$).

\vspace{0.4cm}

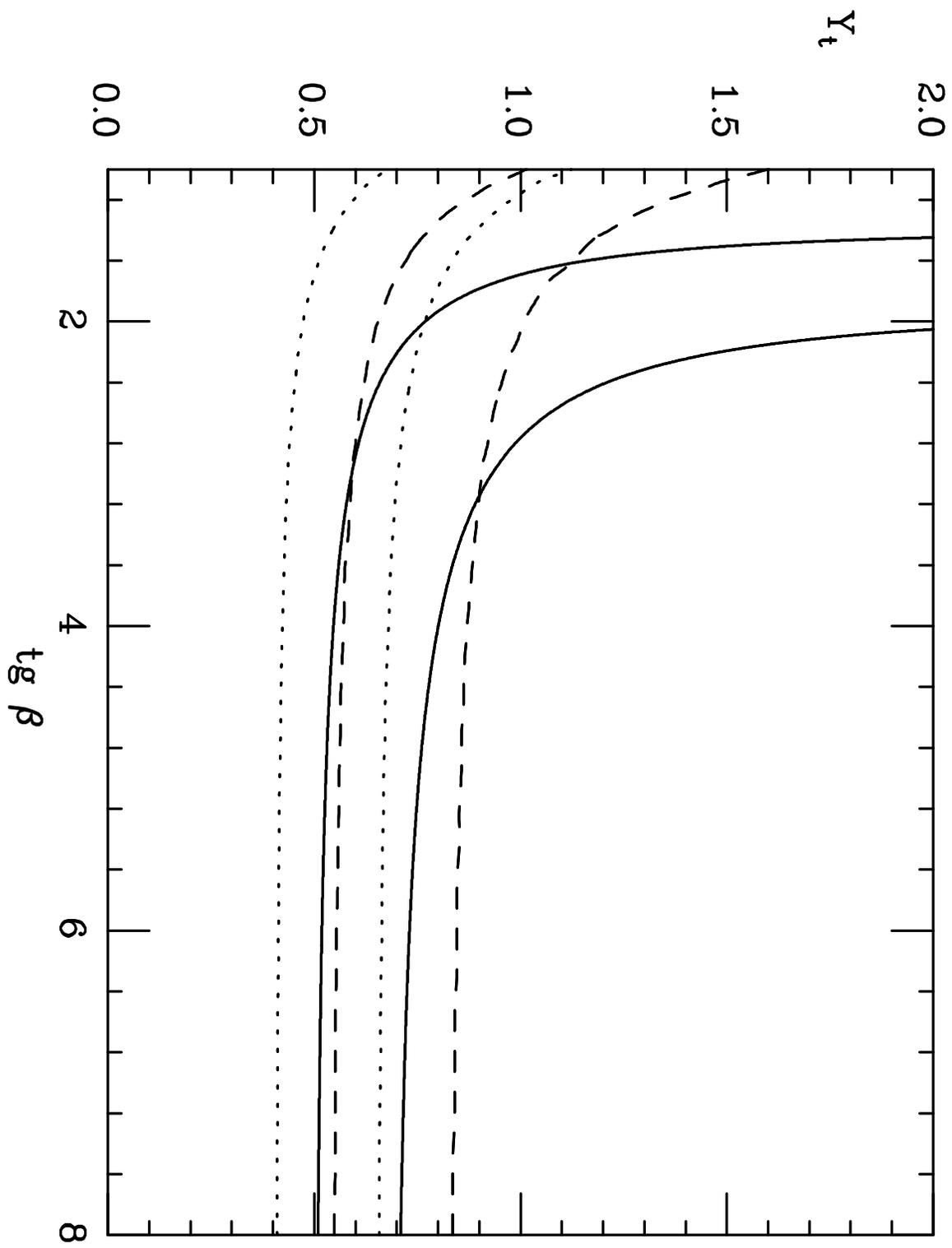
Fig. 5. $~$ The solid isocontours for the top pole-mass 
correspond to 
$M_t= 174$ GeV (upper curve) and $M_t= 160$ GeV (lower curve).  
The dotted iso-contours correspond to 
$s^{q(-)}_{23} = 0.08$ (upper) and $s^{q(-)}_{23} = 0.07$ (lower) 
for the case $b_d=1$ (Fritzsch texture). 
This plot demonstrates that even for the marginal 
values of quark masses used as input ($m_c = 1.5$ GeV,   
$m_b = 4.5$ GeV, $m_s = 100$ MeV), 
the Fritzsch ansatz is completely excluded: 
the magnitude of $|V_{cb}|$ exceeds its experimental value 
at least by factor of 2. 
On the contrary, the case of $b_d=2$ (asymmetric Fritzsch 
texture \cite{ICTP}) can be accomodated:  
the dashed curves are isocontours for $s^{q(-)}_{23} = 0.044$ 
(upper) and $s^{q(-)}_{23} = 0.036$ (lower) and they are 
perfectly compatible with the experimental range of $M_t$.  
Let us remark, however, that $s^{q(-)}_{23}$ represents the 
least possible value of $|V_{cb}|$ (from the destructive interference  
between $s^d_{23}$ and $s^u_{23}$), hence   
for arbitrary phases $\varphi$ the typical values 
of $|V_{cb}|$ are  larger. Clearly, for larger values of 
$b_d$ the allowed range for $\varphi$  becomes also larger. 


\vspace{4mm}

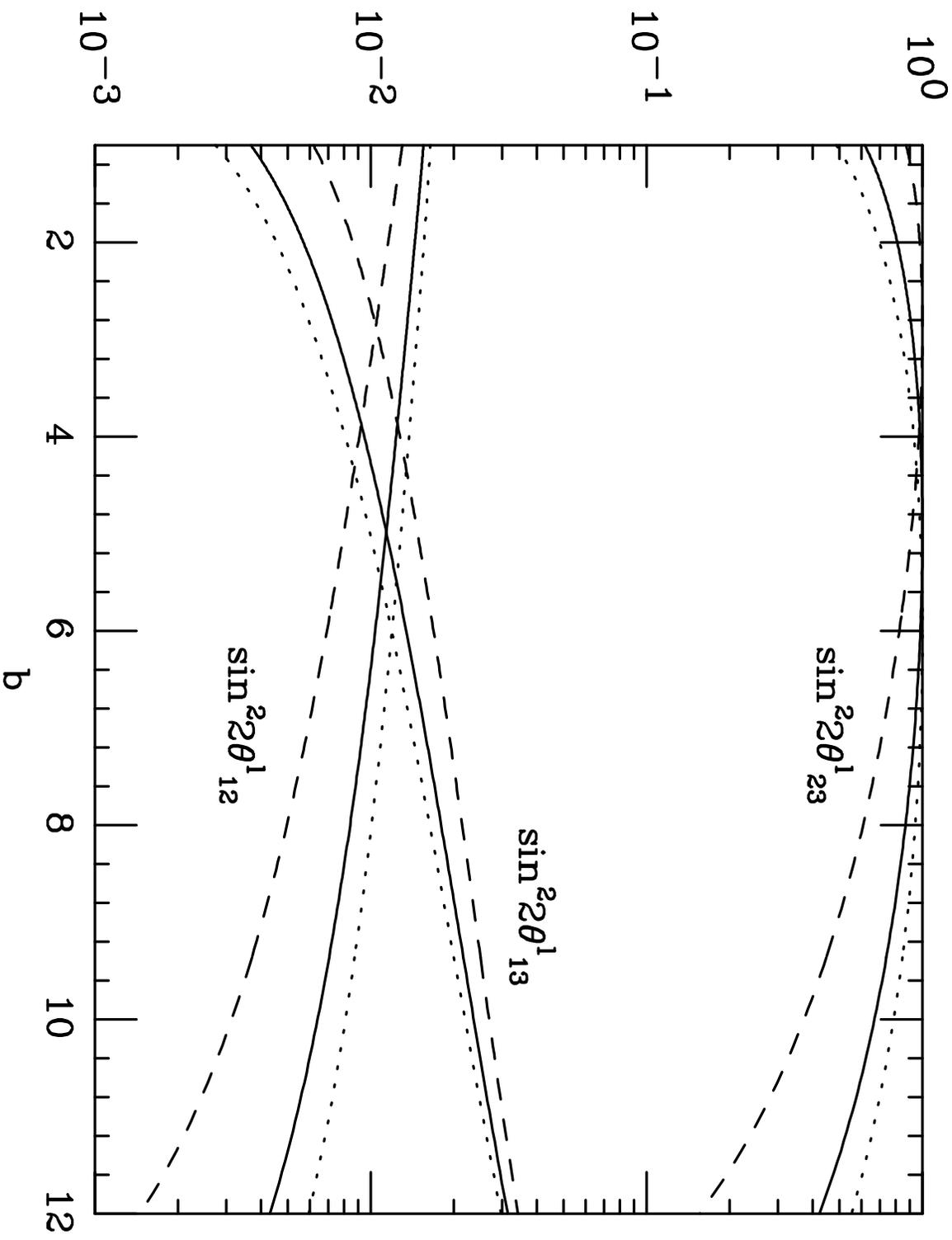
Fig. 6. $~$ The lepton mixing angles as functions of $b$ for 
positive interference between $\theta^e_{23}$ (depending 
on $b$ as in eq. (\ref{ang-23}))  
and $\theta^\nu_{23}=\arctan (\sqrt{m_2/m_3})$. 
The solid line corresponds $\theta^\nu_{23}=12.2^\circ$ 
(central values of $m_{2,3}$ in (\ref{nu-spectr}) ), 
while the marginal possible values 
$20.8^\circ$ (dash) and $8.5^\circ$ (dott) are also shown   
which account for the uncertainties in (\ref{nu-spectr}). 
This plot shows to how small could become 
the parameter $\sin^2 2\theta^l_{12}$ for relevant 
values of $\sin^2 2\theta^l_{23}$.


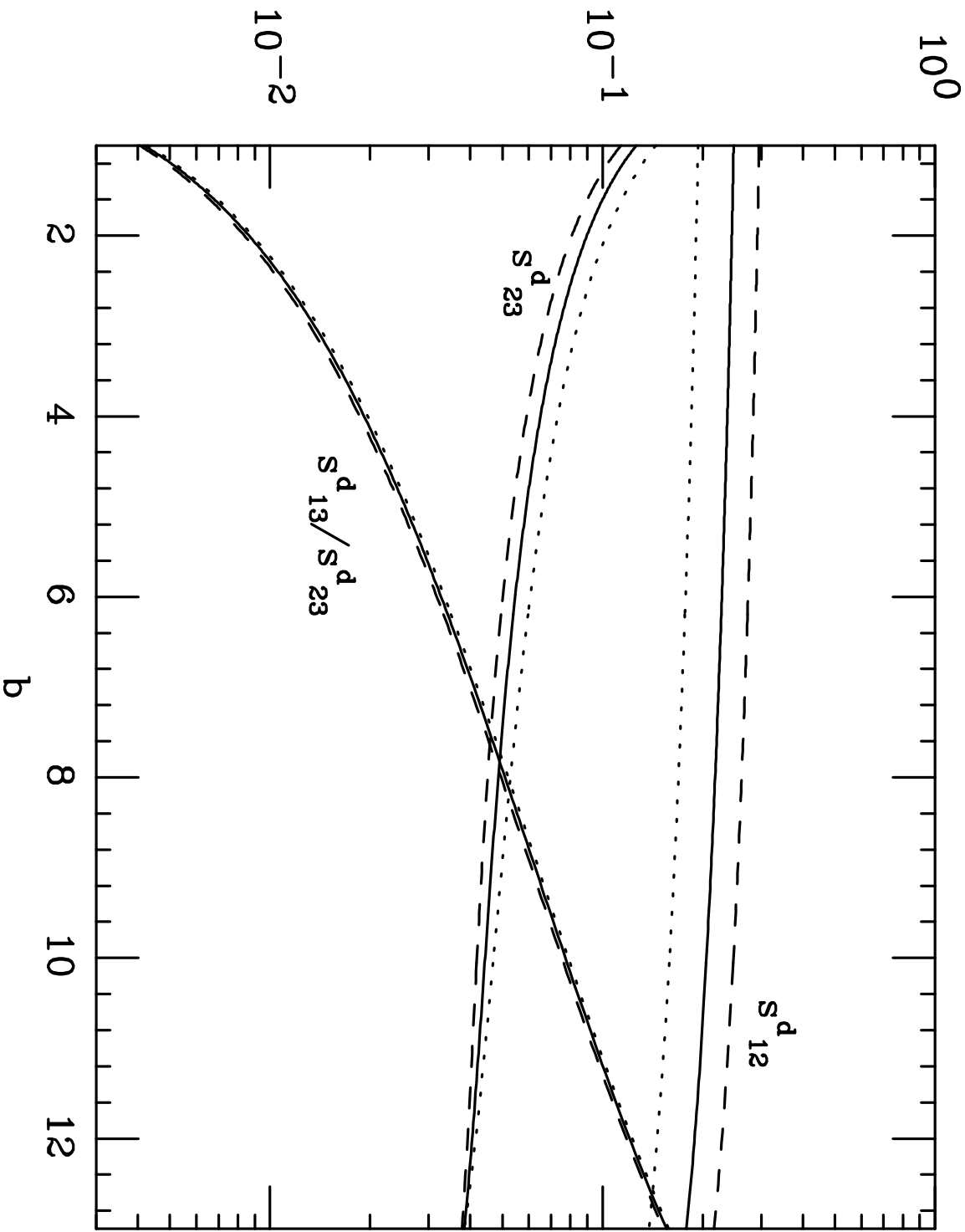
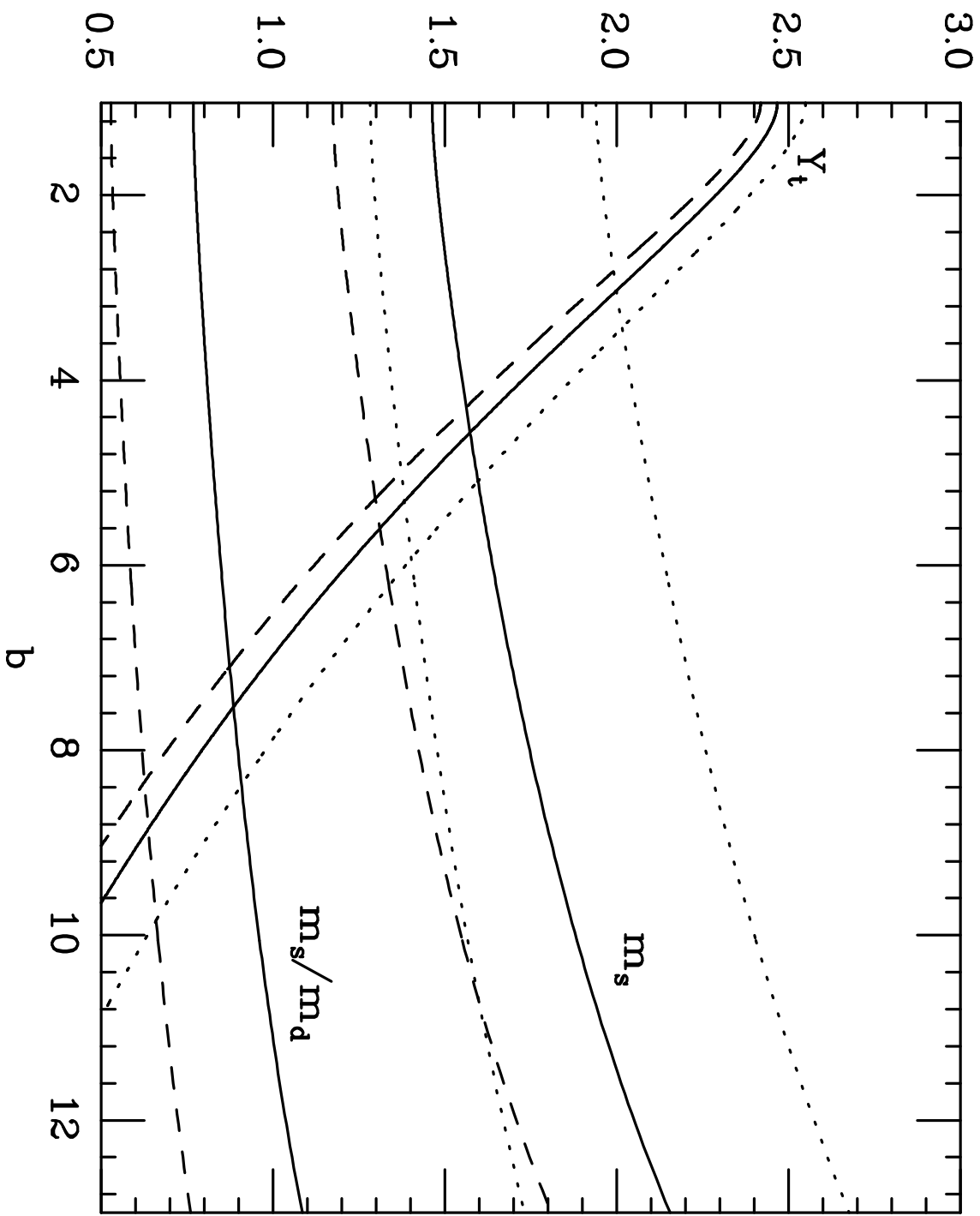
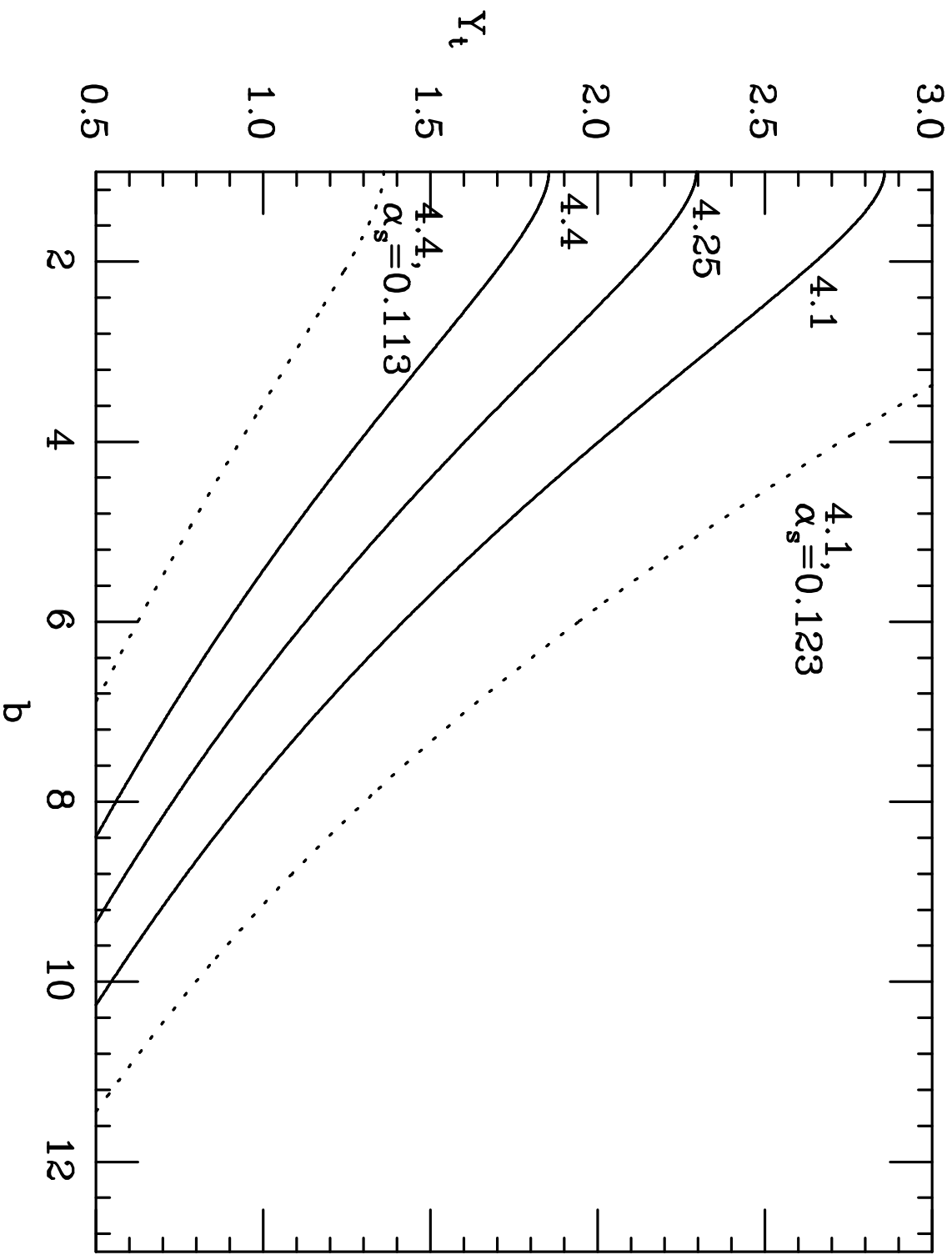
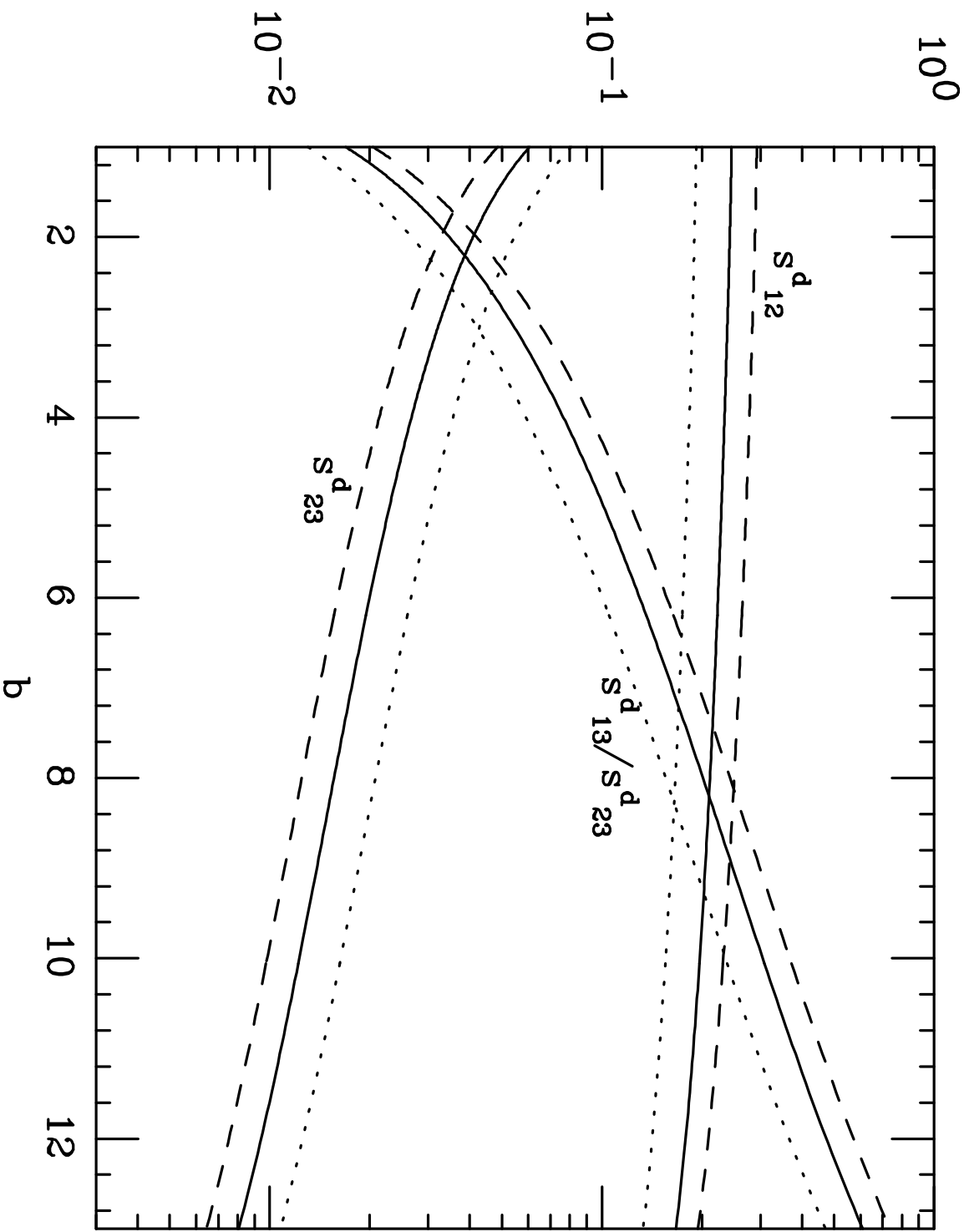
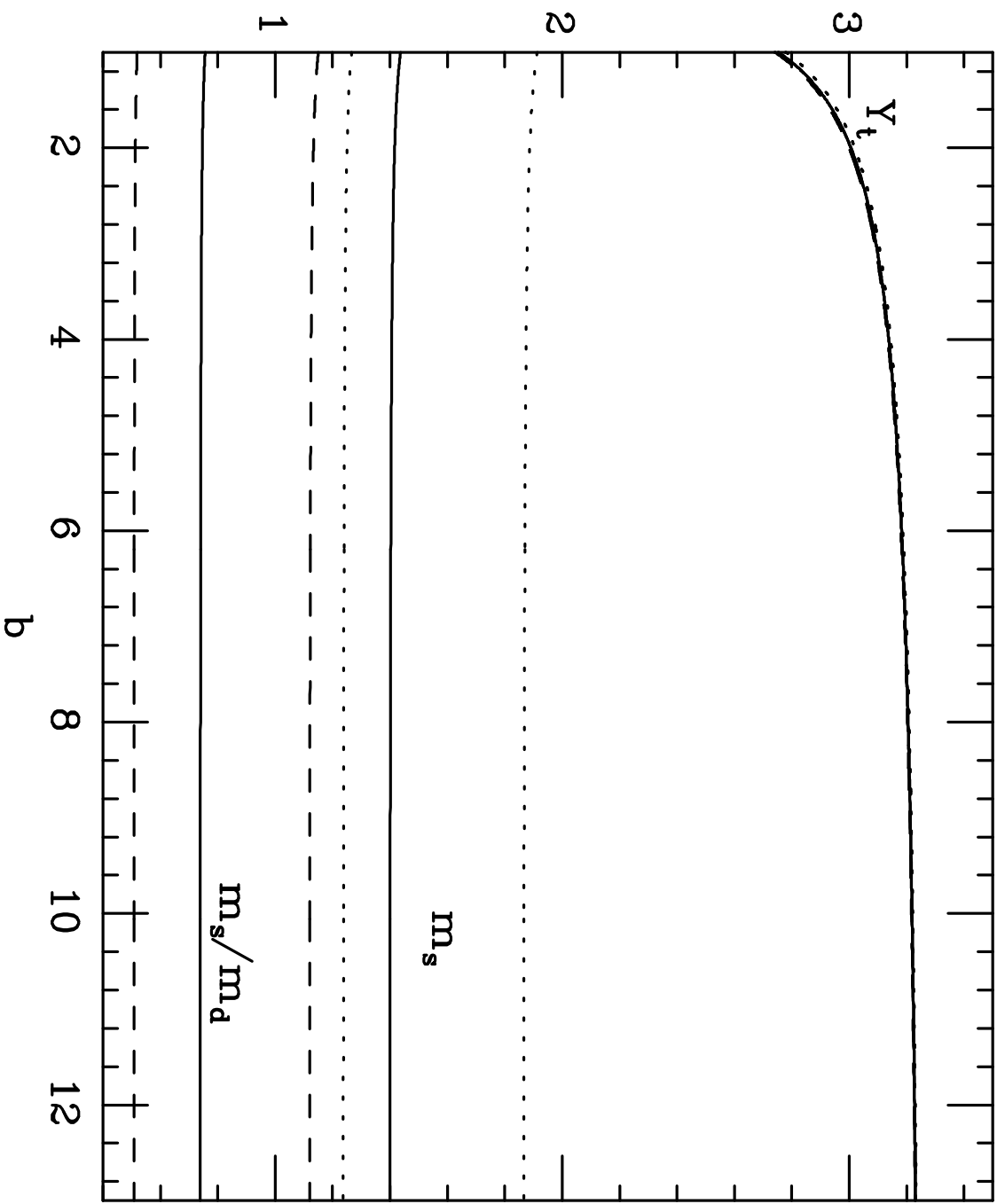
\end{document}